\documentclass[showpacs,preprintnumbers,amsmath,amssymb,prb,twocolumn,floatfix,amssymb,eqsecnum]{revtex4}
\usepackage[dvips]{graphicx,color}
\usepackage{dcolumn,longtable}
\usepackage{amsmath,amssymb,bbold,bm}

\begin{document}
\title{Interplay between Symmetric Exchange Anisotropy, Uniform Dzyaloshinskii-Moriya Interaction and Magnetic Fields in the Phase Diagram of Quantum Magnets and Superconductors}
\author{Ion Garate and Ian Affleck}
\affiliation{Department of Physics and Astronomy, The University of British Columbia, Vancouver, BC V6T 1Z1, Canada, and\\
Canadian Institute for Advanced Research, Toronto, ON M5G 1Z8, Canada.}
\date{\today}

\begin{abstract}
We theoretically study the joint influence of uniform Dzyaloshinskii-Moriya (DM) interactions, symmetric exchange anisotropy (with its axis parallel to the DM vector) and arbitrarily oriented magnetic fields on one-dimensional spin 1/2 antiferromagnets.
We show that the zero-temperature phase diagram contains three competing phases: 
(i) an antiferromagnet with Neel vector in the plane spanned by the DM vector and the magnetic field, 
(ii) a {\em dimerized} antiferromagnet with Neel vector perpendicular to both the DM vector and the magnetic field,  and 
(iii) a gapless Luttinger liquid.
Phase (i) is destroyed by a small magnetic field component along the DM vector and is furthermore unstable beyond a critical value of easy-plane anisotropy, which we estimate using Abelian and non-Abelian bosonization along with perturbative renormalization group.
We propose a mathematical equivalent of the spin model in a one-dimensional Josephson junction (JJ) array located in proximity to a bulk superconductor.
 We discuss the analogues of the magnetic phases in the superconducting context and comment on their experimental viability.
\end{abstract}
\maketitle

\section{Introduction}

Quantum spin chains continue to be an active focus of research largely because they serve as interesting model systems to explore strongly correlated quantum order in low dimensional antiferromagnets,\cite{spin chains} superconductors\cite{jj} and ultracold atoms.\cite{lewenstein2007} 
A significant fraction of current research is devoted to the understanding of frustrated quantum magnets, which display a host of exotic ground states.\cite{proceedings}
One of the agents responsible for magnetic frustration is the Dzyaloshinskii-Moriya\cite{dm} interaction, ${\bf D}_{ij}\cdot({\bf S}_i\times{\bf S}_j)$, which originates from spin-orbit coupling and broken inversion symmetry. 
${\bf S}_i$ is the spin operator at site $i$ and ${\bf D}_{i,j}$ is the DM vector.  
Since this interaction may induce spiral spin arrangements in the ground state,\cite{sudan2009} it is intertwined with ferroelectricity in multiferroic spin chains.\cite{seki2008,huvonen2009}
Besides, the DM interaction plays an important role in explaining the electron spin resonance experiments in some one-dimensional antiferromagnets.\cite{affleck1999}
Moreover, the DM interaction modifies the dynamic properties\cite{derzhko2007} and quantum entanglement\cite{kargarian2009} of spin chains.

The present work is motivated by an elegant recent study\cite{suhas2008} that has predicted intriguing field-induced antiferromagnetic order in Heisenberg spin 1/2 chains with uniform DM interaction (${\bf D}_{ij}=D\hat z$ for any $i,j$).
A crucial aspect of Ref.~[\onlinecite{suhas2008}] is that the external magnetic field is taken to be (nearly) transverse to the DM vector, so that the system has fully broken spin rotational symmetry.
The main objectives of our work are to generalize the analysis of Ref.~[\onlinecite{suhas2008}] for the case of nonzero symmetric exchange anisotropy as well as to find new physical contexts where it might be experimentally testable.

We begin in Section II by introducing the pertinent spin model.
In Section III we identify the classical ground states using the large-spin approximation.
Although some features predicted by the classical analysis are erroneous, most of the classical ground states have a correspondent in the more rigorous quantum analysis performed in Section IV.
In particular, we derive a new result regarding how a magnetic field component along the DM vector modifies the classical soliton-lattice phase. 
 
Section IV is divided in two subsections which complement each other to an extent. 
The first subsection approaches the problem from a non-Abelian bosonization perspective and is constructed around the elegant chiral rotation introduced in Refs.~[\onlinecite{suhas2008,schnyder2008}].  
This subsection constitutes the core of the present work.
The second subsection revisits the problem from an Abelian bosonization viewpoint, and may be skipped on a first reading without loss of continuity.
The phase diagram is richest in the case of weak easy-plane anisotropy.
In this regime there are three competing ground states: 
(i) an antiferromagnet with Neel vector in the plane spanned by the DM vector and the magnetic field, 
(ii) a {\em dimerized} antiferromagnet with Neel vector perpendicular to both the DM vector and the magnetic field,  and 
(iii) a gapless Luttinger liquid.
Phase (i) was first identified in Ref.~[\onlinecite{suhas2008}] at the isotropic exchange point. 
We demonstrate that this phase is fragile under weak-to-moderate easy-plane symmetric exchange anisotropy, and estimate the critical value of the anisotropy beyond which it disappears.
Furthermore, we show that a very small magnetic field component along the direction of the DM vector suffices to destabilize  phase (i) in the neighborhood of the isotropic exchange point.
Thus the experimental detectability of phase (i) appears unlikely except in chains with {\em easy-axis} anisotropy. 
Phases (ii) and (iii) emerge as a consequence of symmetric exchange anisotropy and constitute the main findings of this work.
The Luttinger liquid ground state prevails when the DM interaction is large compared to the magnetic field perpendicular to the DM vector, and the antiferromagnetic phase is stabilized under the opposite condition.
Even though the outcomes of the Abelian and non-Abelian methods agree roughly, the former method misses a few key features such as the coexistence of antiferromagnetism and dimerization in phase (ii).
This is more a merit of the chiral rotation carried out in conjunction with the non-Abelian treatment than an intrinsic flaw of the Abelian bosonization.

In Section V we recast the magnetic model onto a mathematically equivalent problem that consists of a one-dimensional Josephson junction array located in close proximity to a bulk superconductor and placed under a magnetic field.
The analysis of Section IV can be directly transferred to determine the phase diagram of this system at small or large magnetic fields, depending on whether the array is made of $\pi$-junctions or conventional junctions, respectively.
Phases (i) and (ii) discussed in Section IV correspond to a a charge density wave and a vortex lattice, respectively. 
In the vortex lattice phase, circulating supercurrents flow between the array and the bulk superconductor. 
The magnitude of these circulating currents oscillates from one ``plaquette'' to another; this is how dimerization manifests itself in the superconducting context.  
The magnetic Luttinger liquid maps into a state in which the Josephson coupling between the superconducting islands and the bulk superconductor becomes irrelevant.
The transitions between these phases may be controlled with external magnetic fields and by engineering material parameters.
We outline the desiderata for an experimental implementation.

Section VI contains a brief summary of this work and the Appendices include a few technical details concerning the classical and quantum phase diagrams.

\section{Model}

Consider an $S=1/2$ one-dimensional antiferromagnetic chain in presence of a uniform Dzyaloshinskii-Moriya (DM) interaction and an external magnetic field.
Its Hamiltonian is
\begin{eqnarray}
\label{eq:h1}
&&{\cal H}=J\sum_j \left(S^x_j S^x_{j+1} + S^y_j S^y_{j+1} + \Delta S^z_j S^z_{j+1}\right)\nonumber\\
&& +D\hat z\cdot\sum_j \left({\bf S}_j \times {\bf S}_{j+1}\right) - \sum_{j}(h_x S^x + h_z S^z),
\end{eqnarray}
where $J$ is the exchange coupling,  $\Delta$ is the symmetric exchange anisotropy parameter, $D$ is the strength of the DM interaction, $\hat z$ is the direction of the DM vector (chosen to be parallel to the symmetric exchange anisotropy axis), and $h_x$ ($h_z$) is the component of the magnetic field perpendicular (parallel) to the DM vector.   
This model describes one-dimensional magnetic systems with broken inversion symmetry, as well as interacting quantum wires with spin-orbit interactions.\cite{suhas2008}
As we demonstrate in Section V, Eq.~(\ref{eq:h1}) is also germane for a one-dimensional array of Josephson junctions that are proximity coupled to a bulk superconductor (Fig.~\ref{fig:JJ_array}).

Eq.~(\ref{eq:h1}) may be rewritten in a physically more suggestive manner by rotating the spins as $\tilde{S}^+_j=\exp(-i\alpha j) S^+_j$, where $\alpha=\tan^{-1}(D/J)\in[0,\pi/2]$. 
This rotation gauges away the DM interaction and produces a XXZ antiferromagnet with an altered exchange anisotropy $\Delta_{\rm eff}$ and a magnetic field that rotates with a pitch angle $\alpha$ in the plane perpendicular to the DM vector:
\begin{eqnarray}
\label{eq:h2b}
{\cal H}&=&\tilde{J}\sum_j \left( \tilde{S}^x_j \tilde{S}^x_{j+1}+ \tilde{S}^y_j \tilde{S}^y_{j+1} +\Delta_{\rm eff} \tilde{S}^z_j \tilde{S}^z_{j+1}\right)\nonumber\\
&-& h_x \sum_j\left(\tilde{S}^+_j e^{i\alpha j}+\tilde{S}^-_j e^{-i\alpha j}\right)- h_z\sum_j \tilde{S}^z_j,
\end{eqnarray}
where $\tilde{J}=J/\cos\alpha$ and $\Delta_{\rm eff}=\Delta \cos\alpha\leq\Delta$.
When $D\gtrsim J$ the spiral magnetic fields rotates rapidly and thus Eq.~(\ref{eq:h2b}) can be mapped onto a XXZ model with a magnetic field along $\hat z$.
When $D\simeq 0$ the spiral magnetic field rotates very slowly and Eq.~(\ref{eq:h2}) transforms onto a XXZ model with a spatially uniform magnetic field in the $xz$ plane.
In this work we shall be concerned with $0\leq D,h_x,h_z<<J$.

\section{Classical Analysis}

Eq.~(\ref{eq:h1}) is a complicated model with fully broken spin rotational symmetry.
For pedagogical purposes it is useful to begin with simple classical considerations which shed light on the possible ground states of the fully quantum mechanical problem.
When $S$ is large, it is adequate to substitute ${\bf S}_j=S(\sin\theta_j \cos\phi_j,\sin\theta_j\sin\phi_j,\cos\theta_j)$ in Eq.~(\ref{eq:h1}) and seek solutions that satisfy 
$\partial{\cal H}/\partial\phi_i=\partial{\cal H}/\partial\theta_i=0$ for all $i$.
We limit ourselves to $\Delta_{\rm eff}>0$. 
 For $h_z=0$ we find five distinct phases:

(i) Uniform ferromagnet ($FM$), i.e. $\phi_j=0$ (aligned with the field) and $\theta_j=\theta$. There are two solutions: $\theta=\pi/2$ or $\theta=\sin^{-1}(h_x/2 J S (1-\Delta))$. Its energy per site is $\epsilon_{\rm FM}=J S^2 [\Delta+(1-\Delta)\sin^2\theta]-h_x S \sin\theta$.

(ii) Uniform antiferromagnet with Neel vector along $\hat x$ (``$N^x$''), i.e. $\theta_j=\pi/2$ and $\phi_j=\pi j$.
Its energy per particle is $\epsilon_{\rm Nx}=-J S^2$. 

(iii) Uniform antiferromagnet with Neel vector along $\hat y$ (``$N^y$''), i.e. $\theta_j=\pi/2$ and $\phi_j=(-1)^j (\pi/2-\phi_0)$. 
$\phi_0=\sin^{-1}(h_x/4 J S)$ is the canting angle towards the direction of the magnetic field ($\hat x$).
Its energy per site is $\epsilon_{\rm Ny}=-J S^2-h_x^2/8 J$.
$\epsilon_{\rm Ny}<\epsilon_{\rm Nx}$ whenever $h_x\neq 0$.

(iv) Uniform antiferromagnet with Neel vector along $\hat z$ (``$N^z$''), i.e. $\theta_j=(-1)^j\theta_0+\pi j$ and $\phi_j=0$.
  $\theta_0=\sin^{-1}(h_x/2 J S (1+\Delta))$ is the canting angle towards the direction of the magnetic field. 
The energy per site for this phase is $\epsilon_{Nz}=-J S^2 \Delta-h_x^2/4 J (\Delta+1)$.

\begin{figure}
\begin{center}
\scalebox{0.3}{\includegraphics{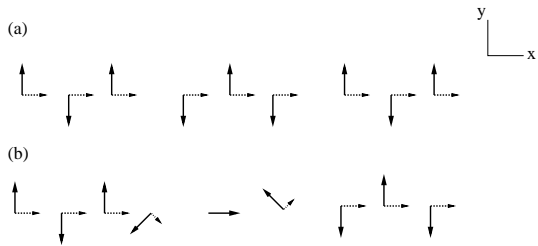}}
\caption{Classical magnetic configuration in the ``$N^y$'' (a) and $LL$ (b) phases when $h_x\neq 0$, $D\neq 0$ and $h_z=0$.
All spins lie in the $xy$ plane. The solid arrows replicate the staggered part of the magnetization whereas the dotted arrows represent the uniform canting towards the direction of the field.
The uniform component of the spins is spatially inhomogeneous in the $LL$ phase and it vanishes at the center of the solitons.
In (b) we limit ourselves to one soliton; for $0<h_x< \pi D S$ these solitons form a periodic array\cite{soliton} with a periodicity that is incommensurate with the underlying lattice.}
\label{fig:soliton}                                                                                                                                                            
\end{center}
\end{figure}

(v) Spiral XY antiferromagnet ($LL$).
For $h_x=0$, this phase is characterized by $\theta_j=\pi/2$ and $\phi_j=\alpha j+\pi j+\chi$.
 $\chi$ is the global angle of the spiral; its arbitrariness renders the $LL$ phase gapless, in contrast to the ones introduced above. 
Its energy per site is $\epsilon_{\rm LL}=-J S^2/\cos\alpha$.
For $h_x\neq 0$, the spiral distorts into an incommensurate soliton lattice\cite{soliton} with $\theta_j=\pi/2$. 
A single soliton is described by $\phi_j=\pi j+a(j)+(-1)^j b(j)$, where $a(j)=2\tan^{-1}[\exp(j h_x/2 \tilde{J} S)]-\pi/2$ and $b(j)=-h_x/2\tilde{J} S \tanh(j h_x/ 2\tilde{J} S)$ vary slowly on the scale of a lattice spacing.
Fig.~\ref{fig:soliton} illustrates the spin arrangement in the soliton.

When $h_x=0$, it is easy to verify analytically that the classical ground state is $LL$ (if $\Delta_{\rm eff}=\Delta\cos\alpha<1$) or $N^z$ (if $\Delta_{\rm eff}>1$).
When $h_x\in(0,\pi D S)$ and $\Delta<1$ the classical ground state is $LL$ (incommensurate soliton lattice).
$h_x=\pi D S$ is the critical field for the commensurate-incommensurate transition.
This critical field is independent of the strength of easy-plane anisotropy because $\theta_j=\pi/2$.
For $h_x\in(\pi D S,4 S J)$ and $\Delta<1$ the ground state is ``$N^y$''. 
When $h_x>4 S J$ the ground state is $FM$. 
For the purposes of this paper $h_x$ will never be large enough to stabilize the $FM$ phase (but see Section V for an exception).
For $\Delta>1$, the ground state is ``$N^z$'' regardless $h_x$.
 
\begin{figure}
\begin{center}
\scalebox{0.3}{\includegraphics{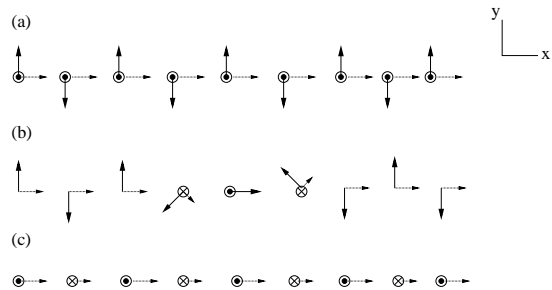}}
\caption{Classical magnetic configuration when $h_x\neq 0$, $D\neq 0$ and $h_z\neq 0$.
(a) ``$N^y$'': the dotted circles indicate uniform canting towards $z$.
(b) $LL$: the spins conform into a soliton lattice in the $xy$ plane much like in Fig.~\ref{fig:soliton}, but they are now canted towards $\hat z$ as well.
As detailed in Appendix A, the latter canting is spatially inhomogeneous. On one hand, it has a non-staggered component (not shown in this figure) that varies slowly along the soliton. In addition, the canting angle has a staggered component (represented in the figure via dotted and crossed circles) that is nonzero only at and near the core of the soliton.   
(c) ``$N^z$'': due to $h_z\neq 0$ the Neel vector is tilted from $z$ towards $x$. Consequently the canting component is not uniform. The crossed circles denote a magnetization component that points into the page.}
\label{fig:hz_phases}                                                                                                                                                            
\end{center}
\end{figure}

Thus far we have neglected $h_z$. 
After turning on $h_z$, ``$N^y$'' develops a uniform canting towards $z$ with $\cos\theta\simeq h_z/2 J S (1+\Delta)$.
On the other hand the soliton develops a canting that has both uniform and staggered components.
As we explain in Appendix A, this leads to a redefinition of the soliton parameters that results in an increased critical field for the commensurate-incommensurate transition: 
\begin{equation}
\label{eq:hc2}
h_{x,c}=\pi \alpha\tilde{J} S \left[1+\frac{h_z^2}{8\tilde{J}^2 S^2 (1-\Delta_{\rm eff}^2)}\right]
\end{equation}
Eq.~(\ref{eq:hc2}) applies when $|\tilde{\Delta}-1|>>h_z^2/\tilde{J}^2 S^2, (d a/ dx -\alpha)^2$.
It also indicates that the influence of $h_z$ on the critical field gets weaker when $\tilde{\Delta}$ decreases.
Regarding ``$N^z$,'' its Neel vector tilts towards $x$ by an angle $\tilde\theta$.
For $\Delta\simeq 1$, $\sin\tilde\theta\simeq -h_z/\sqrt{h_x^2+h_z^2}$ and the Neel vector is nearly perpendicular to ${\bf h}=h_x\hat x+h_z\hat z$. 
Fig.~\ref{fig:hz_phases} illustrates the influence of $h_z$ in the classical ground states.

Turning on $h_z$ modifies the phase diagram qualitatively.
For $\Delta<1$, $LL$ and ``$N^y$'' remain as the ground states, although the critical field for the spin-flop transition is now $\Delta$-dependent.
For $\Delta\gtrsim 1$ the ground state is ``$N^z$'' with a tilted Neel vector.

Some of these classical considerations apply in the case of the $S=1/2$ spin chains, while others do not.
For instance, the occurrence of ``$N^y$'' and gapless (spiral) ground states will be corroborated by the upcoming quantum analysis.
Similarly, our study will confirm that the critical field for the commensurate-incommensurate transition is independent of $\Delta$ at $h_z=0$ but not at $h_z\neq 0$.
In contrast, the quantum analysis will show that ``$N^z$'' can be the ground state even at $\Delta<1$, thus refuting the classical prediction.

\section{Quantum Analysis}

In this section we analyze the exact low energy behavior of Eq.~(\ref{eq:h1}) by combining renormalization group arguments with Abelian and non-Abelian bosonization techniques.
The former technique is best suited for large easy plane anisotropies ($\Delta_{\rm eff}\simeq 0$) whereas the latter method is most reliable at $\Delta_{\rm eff}=1$.
Bearing in mind that each approach has its shortcomings, we shall compare them with each other when possible.
 
\subsection{Non-Abelian bosonization}

Many $S=1/2$ antiferromagnetic chains showcase a weak exchange anisotropy because they involve ${\rm Cu}^{2+}$ compounds.
Furthermore, spin-orbit coupled antiferromagnets without inversion symmetry typically exhibit $\Delta_{\rm eff}\simeq 1$ because the anisotropy induced by the DM interaction nearly cancels the preexisting exchange anisotropy.\cite{anisotropy}

The ground state properties of these nearly isotropic systems may be conveniently accessed using the non-Abelian bosonization.\cite{spin chains,cabra2004,gogolin1998}
In this framework, the spin operators are approximated as
\begin{equation}
{\bf S}_j\to a\left[{\bf J}_L(x)+{\bf J}_R (x) + (-1)^{x/a} {\bf N}(x)\right],
\end{equation}
 where $a$ is the lattice constant. 
${\bf J}_L$ and ${\bf J}_R$ are the uniform components of the left- (right-) spin-currents, respectively.
These currents are conserved at the SU(2) symmetric point ($\Delta_{\rm eff}=1$).
${\bf N}$ is the staggered component of the local spin density.
 
In the continuum limit Eq.~(\ref{eq:h1}) is bosonized in terms of the $SU(2)_1$ Wess-Zumino-Novikov-Witten model.
The low energy effective Hamiltonian can be written in the Sugawara form, which is quadratic in the SU(2) currents: 
\begin{equation}
\label{eq:h_nab}
{\cal H}={\cal H}_0+{\cal H}_{\rm bs}+{\cal V},
\end{equation}
where
\begin{eqnarray}
\label{eq:h_def}
{\cal H}_0&=&\frac{2\pi v}{3}\int dx({\bf J}_R\cdot{\bf J}_R+{\bf J}_L\cdot{\bf J}_L)\nonumber\\
{\cal H}_{\rm bs}&=&-g_{\rm bs}\int dx\left[J^x_R J^x_L+J^y_R J^y_L + (1+\lambda_{\rm xc}) J^z_R J^z_L\right]\nonumber\\
{\cal V} &=& -h_x\int dx (J^x_R + J^x_L)-h_z\int dx (J^z_R + J^z_L)\nonumber\\
&&+\tilde{D}\int dx(J^z_R-J^z_L)-g_{\rm bs}\lambda_{\rm DM}\int dx J^z_R J^z_L,
\end{eqnarray}
 where $v\simeq J a \pi/2$ is the velocity of the low energy excitations near the isotropic point
 (we neglect the anisotropy-induced renormalization of the velocity). 
${\cal H}_0$ is the non-interacting part, the backscattering part ${\cal H}_{\rm bs}$ is the leading marginally irrelevant interaction, and ${\cal V}$ collects the second line of Eq.~(\ref{eq:h1}).
$g_{\rm bs}$ is the (temperature-dependent) coupling constant for the effective interaction, 
$\lambda_{\rm xc}\equiv c(1-\Delta)$ is the symmetric exchange anisotropy parameter and $c$ is a positive constant.
Also, $\tilde{D}\equiv D(1+2\gamma^2)/\pi$ and $\lambda_{\rm DM}=c' D^2/J^2$, where $c'>0$ is another constant.
$\gamma\simeq O(1)$ is the mean-field expectation value of the charge operator.
\begin{equation}
\label{eq:lambda}
\lambda\equiv\lambda_{\rm xc}+\lambda_{\rm DM}
\end{equation}
 is the effective anisotropy parameter, such that $\lambda<0$ corresponds to easy-axis anisotropy (easy axis: $\hat z$) and $\lambda>0$ indicates an easy-plane anisotropy (easy plane: $xy$).

The marginal coupling  $g_{\rm bs}$ at energy scale $T$ was determined accurately from the Bethe ansatz:\cite{lukyanov1998}
\begin{equation}
\label{eq:g_T}
\frac{1}{g(T)}+\frac{1}{2}\log [g(T)]=\log\left[\sqrt{\frac{\pi}{2}}\,e^{\gamma+1/4}\frac{J}{T}\right],
\end{equation}
where $g(T)\equiv g_{\rm bs}/2\pi v$, $\gamma=0.577...$ and $T$ is the temperature.


While deriving the expression for ${\cal V}$ in Eq.~(\ref{eq:h_def}) we have exploited the operator product expansion\cite{gogolin1998} (OPE)
and have kept a higher order term ($\propto D^2/J^2$) when expanding in gradients of $\partial_x\Theta$.
This term has been neglected in previous studies,\cite{suhas2008} yet as we shall show below it is essential in order to reproduce the correct phase diagram in some simple limits.
Hereafter we absorb this term into ${\cal H}_{\rm bs}$.

The constants $c$,$c'$ and $\gamma$ are related to each other and may be determined exactly in certain limits.
For instance, we know that the total anisotropy parameter $\lambda=\lambda_{xc}+\lambda_{\rm DM}$ must vanish when $\Delta_{\rm eff}=1$, i.e. when $\Delta=(1+D^2/J^2)^{1/2}$.
Expanding the square root to leading order in $D/J$, we arrive at $c'=c/2$.
Furthermore, when $h_x=0$ and $\Delta_{\rm eff}<1$ it is well-known (e.g. from the Bethe ansatz method) that the ground state is a Luttinger liquid with a specific value of the Luttinger parameter. 
This determines $c$ in absence of fields (more on this below).
Finally, the value of $\gamma$ may be extracted via the OPE of the DM interaction, which relates $\gamma$ with $c'$. 
For the purposes of this work, the precise value of $\gamma$ will not be important and we shall not be concerned about the distinction between $\tilde{D}$ and $D$ in our numerical calculations.
The renormalization group analysis of Eq.~(\ref{eq:h_nab}) can be simplified considerably by applying a {\em chiral rotation} around the y axis\cite{suhas2008},
which acts differently on right and left currents: 
\begin{eqnarray}
\label{eq:chiral rotation}
{\bf J}_{R(L)}&=&{\cal R}(\theta_{R (L)}) {\bf M}_{R (L)}\nonumber\\
{\cal R}(\theta_{R(L)})&=&\left(\begin{array}{ccc} \cos\theta_{R(L)} & 0 & \sin\theta_{R(L)} \\
                                              0        & 1 &  0\\
                                            -\sin\theta_{R(L)} & 0 & \cos\theta_{R(L)}
\end{array}\right)\nonumber\\
\theta_R &=& \tan^{-1}\left(\frac{d_R}{h_x}\right)-\frac{\pi}{2} \mbox{   ;   } d_R=\tilde{D}-h_z\nonumber\\
\theta_L &=& -\tan^{-1}\left(\frac{d_L}{h_x}\right)-\frac{\pi}{2} \mbox{   ;   } d_L=\tilde{D}+h_z
\end{eqnarray}
The key motivation for the chiral rotation is that it recasts ${\cal V}$ into
\begin{equation}
\label{eq:v_ch}
{\cal V}=\int dx\left( \sqrt{d_R^2+h_x^2} M^z_R+\sqrt{d_L^2+h_x^2} M^z_L\right),
\end{equation}
which is an operator proportional to $M_{R/L}^z$.
\begin{widetext}
Consequently it can be eliminated\cite{giamarchi2003} by the following position-dependent phase shifts:
\begin{eqnarray}
\label{eq:shift}
M^+_R = M^x_R + i M^y_R &\to& M^+_R \exp\left[-i (t_\phi-t_\theta)x\right]\nonumber\\
M^+_L = M^x_L + i M^y_L &\to& M^+_L \exp\left[i(t_\phi+t_\theta)x\right]\nonumber\\ 
M^z_R &\to & M^z_R+\frac{t_\phi-t_\theta}{4\pi}\nonumber\\
M^z_L &\to & M^z_L+\frac{t_\phi+t_\theta}{4\pi},
\end{eqnarray}
 where 
\begin{eqnarray}
t_\phi &=& \left(\sqrt{d_L^2+h_x^2}+\sqrt{d_R^2+h_x^2}\right)/2 v\nonumber\\
t_\theta &=& \left(\sqrt{d_L^2+h_x^2}-\sqrt{d_R^2+h_x^2}\right)/2 v.
\end{eqnarray} 
 The fact that the DM interaction and the magnetic field may be treated on the same footing and absorbed together through phase shifts is a qualitative advantage of the chiral rotation.\cite{schnyder2008}
While ${\cal H}_0$ remains invariant throughout, the successive trasformations modify the backscattering term as follows:
\begin{eqnarray}
\label{eq:h_bs2}
{\cal H}_{\rm bs}&=& 2 \pi v\int dx\left[y_A (M^z_R M^x_L-M^x_R M^z_L) + \tilde{y}_A (M^x_R M^z_L+M^z_R M^x_L)+\sum_{a=x,y,z}y_a M^a_R M^a_L\right]\to {\cal H}_A+\tilde{\cal H}_A+{\cal H}_B+{\cal H}_C+{\cal H}_\sigma\nonumber,
\end{eqnarray}
where
\begin{eqnarray}
{\cal H}_A &=& \pi v\,\, y_A \int dx (M^z_R M^+_L e^{i(t_\phi+t_\theta)x}-M^+_R M^z_L e^{-i(t_\phi-t_\theta)x}+{\rm h.c.})\nonumber\\
\tilde{{\cal H}}_A &=& \pi v\,\, \tilde{y}_A \int dx (M^z_R M^+_L e^{i(t_\phi+t_\theta)x}+M^+_R M^z_L e^{-i(t_\phi-t_\theta)x}+{\rm h.c.})\nonumber\\
{\cal H}_B &=& \pi v\,\, y_B \int dx (M^+_R M^-_L e^{-i 2 t_\phi x}+{\rm h.c.})\nonumber\\
{\cal H}_C &=& \pi v\,\, y_C \int dx (M^+_R M^+_L e^{i 2 t_\theta x}+{\rm h.c.})\nonumber\\
{\cal H}_\sigma &=& -2 \pi v\,\, y_\sigma\int dx M^z_R M^z_L.
\end{eqnarray}
In Eq.~(\ref{eq:h_bs2}) we have neglected small terms that originate from the shifts in $M^z_L$ and $M^z_R$. 
The initial values for the coupling constants in Eq.~(\ref{eq:h_bs2}) are given by
\begin{eqnarray}
\label{eq:init}
y_x(0)&=& -\frac{g_{\rm bs}}{2\pi v}\left[\left( 1+\frac{\lambda}{2}\right)\cos\theta^--\frac{\lambda}{2}\cos\theta^+\right]\nonumber\\
y_y(0) &=& -\frac{g_{\rm bs}}{2\pi v}\nonumber\\
y_z(0)&=&-\frac{g_{\rm bs}}{2\pi v}\left[\left(1+\frac{\lambda}{2}\right)\cos\theta^-+\frac{\lambda}{2}\cos\theta^+\right]\nonumber\\
y_A(0)&=&\frac{g_{\rm bs}}{2\pi v}\left(1+\frac{\lambda}{2}\right)\sin\theta^-\nonumber\\
\tilde{y}_A(0)&=&-\frac{g_{\rm bs}}{2\pi v}\frac{\lambda}{2}\sin\theta^+,
\end{eqnarray}
where $\theta^\pm\equiv \theta_R\pm\theta_L$ and
\begin{eqnarray}
y_C&\equiv&\frac{1}{2}(y_x-y_y)\nonumber\\
y_B &\equiv&\frac{1}{2} (y_x+y_y)\nonumber\\
y_\sigma&\equiv&-y_z.
\end{eqnarray}
\end{widetext}
The anisotropy parameter $\lambda$ modifies the initial couplings and combines with $h_z$ to introduce an extra coupling constant $\tilde{y}_A$ in the RG equations
(note that $\tilde{y}_{A}(0)=0$ when $h_z=0$).

Because of the oscillatory phase factors introduced by Eq.~(\ref{eq:shift}), the RG analysis of ${\cal H}$ must be carried out in multiple stages. 
In the first stage we integrate out momenta that are large compared to ${\rm max}(t_\phi$,$t_\theta$) and thus all phase factors may be ignored in ${\cal H}_{\rm bs}$.
The flow equations can then be derived in the standard manner\cite{gogolin1998} using OPE:
\begin{eqnarray}
\label{eq:rg_nab}
\frac{d y_x}{d l} &=& y_z y_y \nonumber\\
\frac{d y_y}{d l} &=& y_z y_x - (\tilde{y}_A+y_A)(\tilde{y}_A-y_A)\nonumber\\
\frac{d y_z}{d l} &=& y_x y_y\nonumber\\
\frac{d y_A}{d l} &=& y_y y_A\nonumber\\
\frac{d \tilde{y}_A}{d l} &=& -y_y \tilde{y}_A.
\end{eqnarray}
For convenience we begin integrating Eq.~(\ref{eq:rg_nab}) at an initial energy scale $T_0=0.077J$, where the effective coupling has the value $g_{\rm bs}(T_0)\approx 0.23 \times (2\pi v)$ as dictated by Eq.~(\ref{eq:g_T}).          
This is a low enough energy scale, and a small enough value of $g_{\rm bs}(T)$, that the above (lowest order) RG equations apply, at least approximately. 
We then integrate Eq.~(\ref{eq:rg_nab}) towards lower energy scales in order to determine the zero-temperature phase diagram. 
$l\equiv\log(L/a_0)$, where $L$ is the length of the chain and $a_0=v/T_0=20.4 a$ is the ultraviolet RG cutoff lengthscale. 

When $\lambda=0$ Eqs.~(\ref{eq:init}) and ~(\ref{eq:rg_nab}) reduce to those shown in Ref.~[\onlinecite{suhas2008}].
The fact that $\lambda>0$ for an isotropic ($\lambda_{\rm xc}=0$)  Heisenberg antiferromagnet appears to have been overlooked by Ref.~[\onlinecite{suhas2008}], which takes $\lambda_{\rm xc}=0$ and $\Delta_{\rm eff}=1$ simultaneously even in presence of DM interactions.
  
Eq.~(\ref{eq:rg_nab}) is no longer valid when $l>{\rm min}(\log(1/ a_0 t_\phi),\log(1/a_0 t_\theta))$.
For definiteness we assume $t_\theta<<t_\phi$, which holds e.g. when $h_z<<D,h_x$.
Then,  at $l>l_\phi\equiv\log(1/a_0 t_\phi)$, $\exp(i t_\phi x)$ oscillates rapidly and thus the factors multiplying it in ${\cal H}_{\rm bs}$ average to zero.
Therefore $y_A$, $\tilde{y}_A$ and $y_B$ stop renormalizing at $l=l_\phi$. 
In contrast, the factor that multiplies $y_C$ is approximately uniform because $t_\theta<<t_\phi$.
The flow equations for the second RG stage are obtained by setting $y_A=\tilde{y}_A=0$ and $y_B=0$ (i.e. $y_x=-y_y$) in Eq.~(\ref{eq:rg_nab}).
This yields
\begin{eqnarray}
\label{eq:kt_nab}
\frac{d y_C}{d l} &=& y_\sigma y_C \nonumber\\
\frac{d y_\sigma}{d l} &=& y_C^2, 
\end{eqnarray}
which are the famous Kosterlitz-Thouless (KT) equations with known analytic solution.
The ``initial'' conditions for the second RG stage are $y_\sigma(l_\phi)=-y_z(l_\phi)$ and $y_C(l_\phi)=(y_x(l_\phi)-y_y(l_\phi))/2$.
We integrate Eq.~(\ref{eq:kt_nab}) from $l=l_\phi$ up until $l=l_\theta\equiv\log(1/a_0 t_\theta)$.
At $l=l_\theta$ the flow of $y_C$ stops because its coefficient in ${\cal H}_{\rm bs}$ contains a $\exp(i t_\theta x)$ factor.
By setting $y_C=0$ in Eq.~(\ref{eq:kt_nab}), it follows that $y_\sigma$ stops flowing as well.
Thus at $l=l_\theta$ the system is in its ground state, the nature of which is determined by the final values of the coupling constants.

In order to elicit the physical meaning  of the different ground states we resort to the relation\cite{gogolin1998} between the non-Abelian operators and the bosonic fields ($\tilde\Theta$, $\tilde\Phi$) with which the Abelian bosonization is constructed:
\begin{eqnarray}
\label{eq:equiv}
M^+_R &=& \frac{1}{2\pi a} e^{-i\sqrt{2\pi} (\tilde\Phi-\tilde\Theta)}\mbox{   ;   } M^+_L = \frac{1}{2\pi a} e^{i\sqrt{2\pi} (\tilde\Phi+\tilde\Theta)}\nonumber\\
M^z_L &=& \frac{1}{2\sqrt{2\pi}}(\partial_x\tilde\Phi+\partial_x\tilde\Theta)\mbox{    ;    } M^z_R = \frac{1}{2\sqrt{2\pi}}(\partial_x\tilde\Phi-\partial_x\tilde\Theta)\nonumber\\
{\cal N}^\pm &=& \frac{\gamma}{\pi a} e^{\pm i\sqrt{2\pi}\tilde\Theta}\nonumber\\
{\cal N}^z &=& \frac{\gamma}{\pi a} \sin(\sqrt{2\pi}\tilde\Phi),
\end{eqnarray}
where $({\cal N}^x,{\cal N}^y,{\cal N}^z)$ is the staggered magnetization in the rotated frame.
$\tilde\Phi$ and $\partial_x\tilde\Theta$ are canonically conjugate fields, i.e. $[\tilde\Phi (x),\partial_x\tilde\Theta (x')]=i\delta(x-x')$.
We reserve the $\Theta$ and $\Phi$ notation (without tilde) for the next subsection, where we shall employ Abelian bosonization in a different coordinate system.
Eq.~(\ref{eq:equiv}) yields $M^+_R M^+_L +{\rm h.c.}\propto \cos(\sqrt{8\pi}\tilde\Theta)$ and  $M^+_R M^-_L +{\rm h.c.} \propto\cos(\sqrt{8\pi}\tilde\Phi)$.
Hence if $y_C(l_\theta)\to\pm\infty$ the minimum energy state corresponds to $\cos(\sqrt{8\pi}\tilde\Theta)=\mp 1$, which implies that $\tilde\Theta$ gets ordered and a gap is opened in the spin excitation spectrum.
Furthermore, Eq.~(\ref{eq:equiv}) shows that when $\cos(\sqrt{8\pi}\tilde\Theta)=1 (-1)$ there is long-range antiferromagnetic order with the Neel vector pointing along $\hat x$ ($\hat y$) in the {\em rotated} frame.

When $|y_B(l_\theta)|\to\infty$ the minimum energy state corresponds to $\cos(\sqrt{8\pi}\tilde\Phi)=\mp 1$.
Once again resorting to Eq.~(\ref{eq:equiv}), it follows that when $\cos(\sqrt{8\pi}\tilde\Phi)=1 {\rm or } -1$ the system settles into a dimerized state or an antiferromagnetic state with Neel vector along $\hat z$ in the {\em rotated} frame, respectively.
Both phases are gapped because $\tilde\Phi$ is ordered.

In the $LL$ phase neither $\tilde\Theta$ nor $\tilde\Phi$ order.
This gapless Luttinger Liquid has dominant dimer or spin density wave correlations depending on whether $y_C(l_\theta)$ dominates over $y_B(l_\theta)$ or vice versa.
At first glance a gapless ground state appears unlikely in a model with completely broken spin rotational symmetry; however, it is not unprecedented.
Similar states with gapless spin excitation spectra occur in {\em magnetized}, spin-orbit coupled, one-dimensional conductors that are placed under a magnetic field.\cite{giamarchi1988} 
As we shall see below, the $LL$ ground state results when $D$ dominates over $h_x$. 

In order to determine what the aforementioned ordered states mean in terms of the original spin variables ${\bf J}$ and ${\bf N}$, we once again perform the rotation introduced in Eq. ~(\ref{eq:chiral rotation}) and arrive at
\begin{eqnarray}
\label{eq:op_transf}
N^x &=&\cos\left(\frac{\theta^+}{2}\right) {\cal N}^x - \sin\left(\frac{\theta^+}{2}\right) {\cal N}^z\nonumber\\
N^z &=& \sin\left(\frac{\theta^+}{2}\right) {\cal N}^x + \cos\left(\frac{\theta^+}{2}\right) {\cal N}^z\nonumber\\
N^y &=& \cos\left(\frac{\theta^-}{2}\right) {\cal N}^y - \sin\left(\frac{\theta^-}{2}\right){\cal E}\nonumber\\
\epsilon &=& \sin\left(\frac{\theta^-}{2}\right) {\cal N}^y + \cos\left(\frac{\theta^-}{2}\right){\cal E}.
\end{eqnarray}
${\cal N}^i$ and ${\cal E}$ denote the Neel vectors and the dimerization in the rotated frame, respectively.
$N^i$ and $\epsilon$ are their counterparts in the original coordinate system.
The derivation of Eq.~(\ref{eq:op_transf}) becomes straightforward after recognizing that\cite{cabra2004}
\begin{eqnarray}
{\bf M}_R &\propto& {\rm tr}[{\boldsymbol\sigma} g^{-1} \partial_z g] \mbox{    ;    } {\bf M}_L \propto {\rm tr}[{\boldsymbol\sigma} g^{-1} \partial_{\bar z} g]\nonumber\\
{\bf{\cal N}} &=& {\rm tr}[{\boldsymbol\sigma} g]\nonumber\\
{\cal E} &=& {\rm tr}[g]
\end{eqnarray}
where ${\boldsymbol\sigma}$ is a vector of Pauli matrices, $\partial_{z(\bar z)}=\partial_t/v+(-)\partial_x$ and
\begin{eqnarray}
g\propto\left(\begin{array}{cc} e^{i\sqrt{2\pi}\Phi} & e^{-i\sqrt{2\pi}\Theta}\\
                          -e^{i\sqrt{2\pi}\Theta} & e^{-i\sqrt{2\pi}\Phi}
\end{array}\right) 
\end{eqnarray}
is the SU(2) matrix field that enters the WZNW action and transforms as
\begin{equation}
g\to e^{i\sigma^y\theta_L/2} g e^{-i\sigma^y\theta_R/2}
\end{equation}
under a chiral rotation (this $g$ is of course not to be confused with its homonym of Eq.~(\ref{eq:g_T})) .
While we find $\langle{\cal E}\rangle = 0$ for all values of $D, h_x$ and $h_z$, there exist regions of parameter space for which $\langle{\cal N}^y\rangle\neq 0$. 
This then translates into a coexistence of antiferromagnetism (with Neel vector along $\hat y$) and dimerization in the original frame.
Table 1 enumerates and characterizes this and other possible ground states of Eq.~(\ref{eq:h1}) in the original coordinate system.

\begin{table*}[t]
\caption{Ground states for Eq.~(\ref{eq:h1}), based on the value of the coupling constants at the end of the RG flow.
Our calculations show that $\langle{\cal E}\rangle=0$ for any $h_x,h_z$ and $D$.
However, this does not preclude a dimerized phase in Eq.~(\ref{eq:h1}) because the gapped phase labeled as ``$N^y$" contains a mixture of Neel-y correlations along with dimerization.
Our RG analysis demonstrates that ``$N^y$" emerges when $D<<h_x$ and $\Delta<1$.
It follows from Eq.~(\ref{eq:op_transf}) that the dimerization component of this phase is enhanced as $D$ gets larger.   
The nomenclature for gapped phases labeled as ``$N^x$" and ``$N^z$" is motivated by the $h_z=0$ case, for which ``$N^x$"(``$N^z$") means antiferromagnetic order with Neel vector along $\hat x$ ($\hat z$).
When $h_z\neq 0$ the Neel vector lies in the $xz$ plane for both phases.}
\vspace{0.1 in}
\begin{tabular}{c c c c c}
\hline\hline
$y_B(l_\theta)$ & & $y_C(l_\theta)$ & Ground State & Ground State\\
 & & & (Rotated Frame) & (Original Frame)\\
\hline\\
$+\infty$ &  & finite  &   $\langle{\cal E}\rangle\neq 0$ & ``$\epsilon$'': $\langle\epsilon\rangle\neq 0\neq\langle N^y\rangle$ 
\\[1ex]
\hline\\
$-\infty$ & & finite & $\langle{\cal N}^z\rangle \neq 0$ & ``$N^x$'': $\langle N^x\rangle\neq 0\neq\langle N^z\rangle$
\\[1ex]
\hline\\
finite & & $-\infty$ & $\langle{\cal N}^y\rangle \neq 0$ & ``$N^y$'': $\langle\epsilon\rangle\neq 0\neq\langle N^y\rangle$ 
\\[1ex]
\hline\\
finite & & $+\infty$ & $\langle{\cal N}^x\rangle \neq 0$ & ``$N^z$'': $\langle N^x\rangle\neq 0\neq \langle N^z\rangle$
\\[1ex]
\hline\\
finite & & finite & disordered & disordered
\\[1ex]
\hline
\end{tabular}
\end{table*}

In Appendix B we digress on some simple limits in which the phase diagram is known with certainty.
Besides providing a reality check, this enables us to determine the value of the constant $c$ introduced above Eq.~(\ref{eq:lambda}).
This appendix may be skipped on a first reading.
  
\begin{figure}
\begin{center}
\scalebox{0.3}{\includegraphics{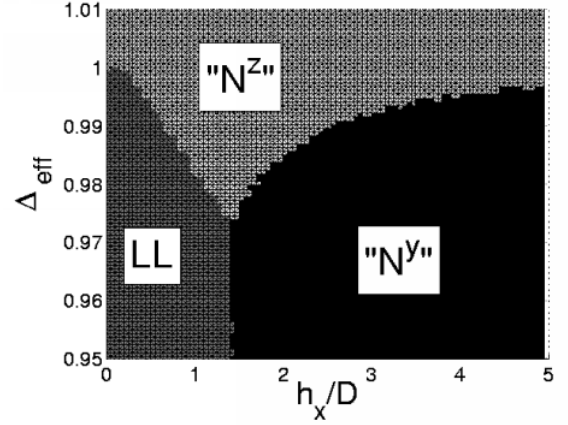}}
\caption{Phase diagram corresponding to Eq.~(\ref{eq:h1}); derived using non-Abelian bosonization. 
We fix $D=5\times10^{-4} J$ and $h_z=0$; we vary $h_x$ and $\Delta_{\rm eff}$ (note that $\Delta_{\rm eff}$ and $\Delta$ are nearly identical in this figure). 
For $\Delta_{\rm eff}>1$ (easy axis anisotropy), the ground state is unequivocally ``$N^z$''. 
For $\Delta_{\rm eff}<1$ (easy plane anisotropy), two new phases emerge: the gapless $LL$ at $D>>h_x$  and the gapped ``$N^y$'' at $D<<h_x$. 
The phase line separating ``$N^y$'' and ``$N^z$'' has a horizontal asymptote at $\Delta_{\rm eff}=1$ as $h_x/D\to\infty$.
Below a critical value of the anisotropy $(\Delta_{\rm eff}<\Delta_c\simeq 0.973$ in this figure) ``$N^z$'' can no longer be the ground state.} 
\label{fig:fig1}                                                                                                                                                               
\end{center}
\end{figure}

We now embark on the numerical study of the general phase diagram for Eq.~(\ref{eq:h1}).
Fig.~\ref{fig:fig1} characterizes the influence of the effective exchange anisotropy for $h_x\neq 0$ and $h_z=0$.
When $\Delta_{\rm eff}>1$ ($\lambda\leq 0$) the only available ground state is ``$N^z$''.
When $\Delta_{\rm eff}<1$ ($\lambda>0$), ``$N^y$'' prevails at $h_x/D>>1$ and $LL$ reigns at $h_x/D<<1$.
From Eq.~(\ref{eq:op_transf}) it is clear that ``$N^y$'' contains dimerization that is most noticeable near the phase boundary with $LL$, fading away as $h_x/D\to\infty$. 
As the easy-plane anisotropy gets stronger, the range of $h_x/D$ for which ``$N^z$'' is the ground state becomes narrower.
 As a matter of fact ``$N^z$'' disappears completely for $\Delta_{\rm eff}<\Delta_c$, where
the critical value $\Delta_c$ depends on $h_x,D$ and $h_z$.
We shall revisit and reafirm this point in the next subsection.
For $\Delta_{\rm eff}<\Delta_c$, the phase boundary between $LL$ and ``$N^y$'' is independent of $\Delta$ and occurs at $h_x\simeq 1.5 D$.
This regime matches fairly well with the classical predictions of Section III, where the critical field for the commensurate-incommensurate transition was found to be $h_{x,c}=\pi D S=\pi D/2$. 

\begin{figure}
\begin{center}
\scalebox{0.3}{\includegraphics{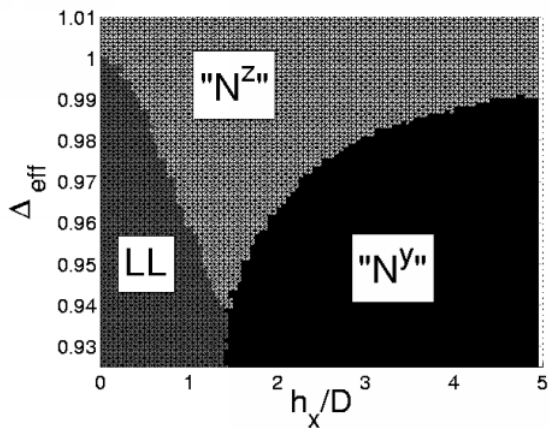}}
\caption{Phase diagram corresponding to Eq.~(\ref{eq:h1}); derived using non-Abelian bosonization. 
$D=0.01 J$ and $h_z=0$; we vary $h_x$ and $\Delta$. 
Comparing this plot with Fig.~\ref{fig:fig1} it is clear that for a given $D/h_x$, ``$N^z$'' is more robust when $D$ is larger. 
In other words, $\Delta_c$ ($\simeq 0.94$ in this figure) decreases as $D$ and $h_x$ increase.}           
\label{fig:fig2}                                                                                                                                                               
\end{center}
\end{figure}

Fig.~\ref{fig:fig2} differs from Fig.~\ref{fig:fig1} only quantitatively, but serves to highlight that ``$N^z$'' is more robust for larger values of $D$ and $h_x$ even as $h_x/D$ is unchanged.
This observation can be understood as follows.
When $\Delta_{\rm eff}<1$, 
the value of $y_C$ decreases during the first stage of RG.
When $D$ and $h_x$ are very small, $l_\phi$ is large and the prerequisite for flowing to ``$N^z$'' ($y_C(l_\phi)>-y_\sigma(l_\phi$) at $y_\sigma<0$) is less likely to be fulfilled.
The larger $D$ and $h_x$ are, the smaller $l_\phi$ and thus the shorter the decay of $y_C$; this improves the odds for a ``$N^z$'' ground state. 

\begin{figure}[h]
\begin{center}
\scalebox{0.3}{\includegraphics{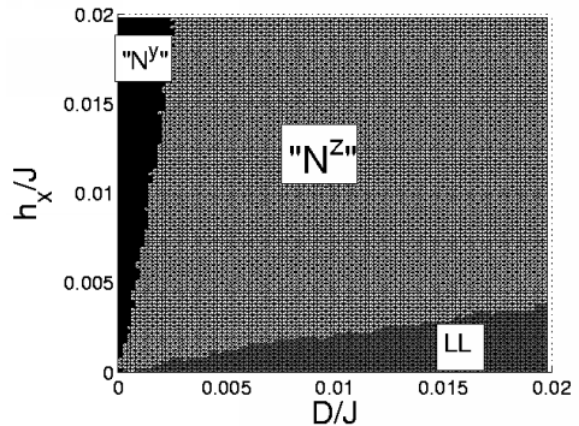}}  
\caption{Phase diagram with $\lambda\simeq c(1-\Delta_{\rm eff})\simeq 0.02$ and $h_z=0$; derived using non-Abelian bosonization. ``$N^y$'' prevails when $D/h\gtrsim\sqrt{\lambda}$ and $LL$ prevails when $h/D\lesssim\sqrt{\lambda}$. These results may be understood analytically, as discussed in the text.}                       
\label{fig:fig4}                                                        
\end{center}
\end{figure}

Fig.~\ref{fig:fig4} shows that for a given anisotropy the phase boundaries between $LL$, ``$N^y$'' and ``$N^z$'' are linear.
It is straightforward to explain this behavior analytically.
Since $h_z=0$ we have $\theta^+=-\pi$ and hence $\tilde{y}_A(l)=0$.
Furthermore, we shall assume that $y_A$ is small in the first stage of the RG flow.
This assumption is adequate only if $\sin\theta^-\simeq 0$.
Fortunately, this condition is satisfied when $D>>h_x$ ($\theta^-\simeq\pi$) or $D<<h_x$ ($\theta^-\simeq 0$), which are the regions of interest when seeking ``$N^y$'' or $LL$.
Thus neglecting $y_A$ and $\tilde{y}_{A}$ from the onset, we are left with KT equations for $y_C$ and $y_\sigma$, the initial conditions being
\begin{eqnarray}
y_\sigma(0) &=& \frac{g_{\rm bs}}{2\pi v}\left[\left(1+\frac{\lambda}{2}\right)\cos\theta^--\frac{\lambda}{2}\right]\nonumber\\
y_C(0) &=& -\frac{g_{\rm bs}}{4\pi v}\left[\left(1+\frac{\lambda}{2}\right)\cos\theta^--1+\frac{\lambda}{2}\right]
\end{eqnarray}
On one hand, the KT equations lead to ``$N^y$'' provided that $y_C<0$ and $y_\sigma>0$ (or alternatively if $y_C<y_\sigma$ and $y_\sigma<0$, although this is not satisfied at small $\lambda$).
After some algebra this amounts to requesting $\theta^-\in (-2\sqrt{\lambda},2\sqrt{\lambda})$, for $\lambda<<1$. 
Using $D/h_x=\tan\theta^-\simeq\theta^-$, it follows that ``$N^y$'' exists when $0<D/h_x<\sqrt{\lambda}$. 
This is why we have a linear phase boundary between ``$N^y$'' and $LL$.
Moreover, the slope of the corresponding line in Fig.~\ref{fig:fig4} matches fairly well with $\sqrt{\lambda}$.
On the other hand, the KT equations predict $LL$ if $y_\sigma<0$ and $y_C<y_\sigma<-y_C$.
These inequalities may be reduced to $\cos\theta^-<2\lambda-1$, where we have expanded for small $\lambda$.
This condition is satisfied for $\theta^-\in(\pi-2\sqrt{\lambda},\pi)$ or equivalently $D/h_x\gtrsim\sqrt{\lambda}$, which is another straight line. 

\begin{figure}[h]
\begin{center}
\includegraphics[scale=0.3, angle=-90]{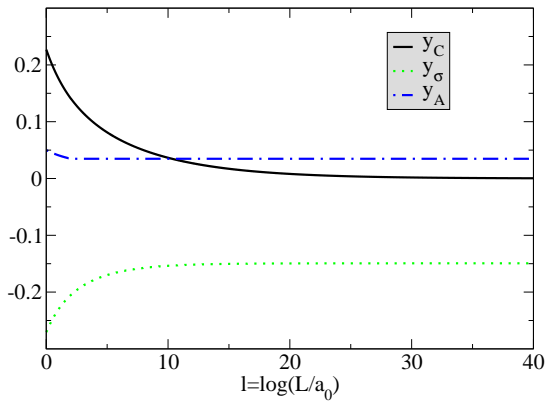}
\caption{\label{fig:frg1} (color online) Typical flow of the coupling constants in the $LL$ phase. $D=0.01 J$, $h_x/D=0.1$, $h_z=0$ and $\Delta\simeq 0.975$. $\tilde{y}_{A}(l)=0$ and $y_B (l)\simeq 0$ (not shown).
A horizontal line indicates that a particular coupling constant has stopped flowing at because of rapid spatial oscillations. $l_\phi\simeq 2$, $l_\theta=\infty$.}
\end{center}
\end{figure}

\begin{figure}[h]
\begin{center}
\includegraphics[scale=0.3, angle=-90]{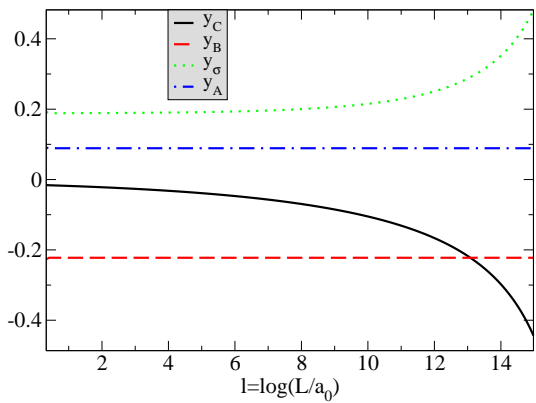}
\caption{\label{fig:frg2} (color online) Typical flow of the coupling constants in the ``$N^y$'' phase. $D=0.01 J$, $h_x/D=5$, $h_z=0$ and $\Delta\simeq 0.975$. $\tilde{y}_{A}(l)=0$(not shown). $l_\phi\simeq 0.4$, $l_\theta=\infty$.}
\end{center}
\end{figure}

\begin{figure}
\begin{center}
\includegraphics[scale=0.3, angle=-90]{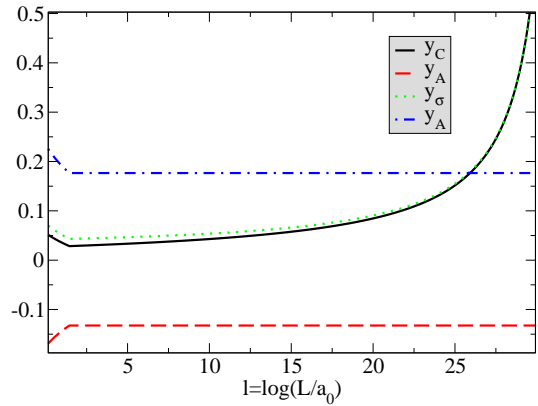}
\caption{\label{fig:frg3} (color online) Typical flow of the coupling constants in the ``$N^z$'' phase. $D=0.01 J$, $h_x/D=1.5$, $h_z=0$ and $\Delta\simeq 0.975$. $\tilde{y}_{A}(l)=0$(not shown). $l_\phi\simeq 1.5$, $l_\theta=\infty$.
Had we chosen a smaller $\Delta$ the strong coupling would have been reached at larger value of $l$.
This is why the ``$N^z$'' phase at $\Delta<1$ is particularly fragile to $h_z\neq 0$.}
\end{center}
\end{figure}

For completeness we include Figs.~\ref{fig:frg1},~\ref{fig:frg2} and~\ref{fig:frg3}, which display typical RG flows for the coupling constants in each of the phases.

\begin{figure}
\begin{center}
\includegraphics[scale=0.3]{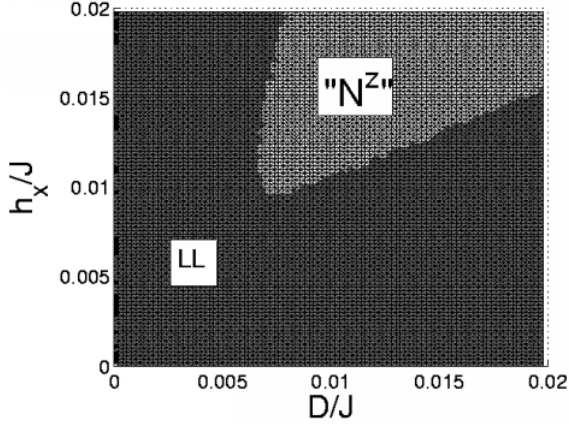}
\caption{\label{fig:fig5} Influence of $h_z$ on the phase diagram of Eq.~(\ref{eq:h1}); derived using non-Abelian bosonization. $h_z=10^{-6} h_x$ and $\lambda\simeq c (1-\Delta_{\rm eff})\simeq 0.02$. 
The $LL$ ground state is defined (somewhat arbitrarily) via $|y_C(l_\theta)|<0.2$. Even for a small $z$-component of the field the outcome is dramatically different from Fig.~\ref{fig:fig4}. In this figure ``$N^y$'' is no longer present (it resurfaces for smaller $\Delta_{\rm eff}$) and ``$N^z$'' can be found only in a limited region of the parameter space (that of $D\simeq h_x$ and relatively large $D$). The reason why ``$N^y$'' is more fragile than ``$N^z$'' is that the value of $l$ at which it reaches strong coupling is larger. As we shall see in the next subsection, this trend reverses when the easy-plane anisotropy is stronger; in that case ``$N^z$'' is more fragile than ``$N^y$''.} 
\end{center}
\end{figure}

\begin{figure}
\begin{center}
\includegraphics[scale=0.3]{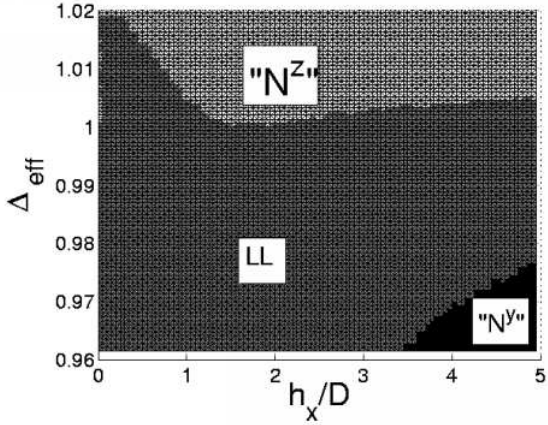} 
\caption{\label{fig:fig_hz}Influence of $h_z$ on the phase diagram of Eq.~(\ref{eq:h1}); derived using non-Abelian bosonization. $h_z=10^{-5} h_x$, $D=0.01 J$. 
The $LL$ ground state is defined (somewhat arbitrarily) via $|y_C(l_\theta)|<0.2$. 
Comparing this figure with Fig.~\ref{fig:fig2}, it is clear that even a small $h_z$ brings about qualitative changes to the phase diagram.
Remarkably, not only ``$N^z$'' is no longer the ground state for $\Delta<1$, but even for $\Delta>1$ there is a swath of parameter space for which $LL$ prevails.
For $\Delta_{\rm eff}<1$ the phase boundary between $LL$ and ``$N^y$'' is pushed to larger values of $h_x$. This is in qualitative agreement with Eq.~(\ref{eq:hc2}): the critical field for the commensurate-incommensurate transition increases due to $h_z$, such increase being less pronounced as $\Delta$ is made smaller.}
\end{center}
\end{figure}

\begin{figure}
\begin{center}
\includegraphics[scale=0.3, angle=-90]{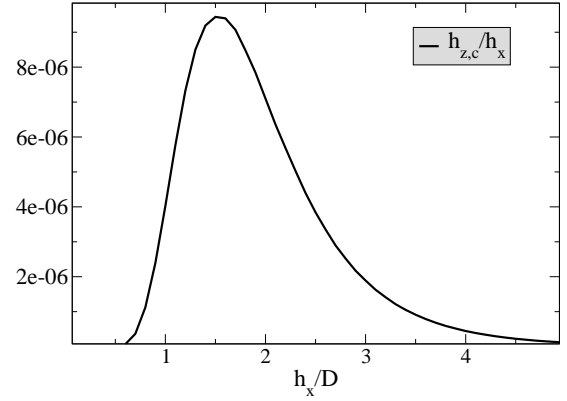}
\caption{\label{fig:hzc} Plot of Eq.~(\ref{eq:hzc}) for $\Delta=1$ and $D\simeq 0.1 J$. When $h_z>h_{\rm c,z}$ ``$N^z$'' is no longer the ground state. $h_{\rm z,c}$ is largest when $h_x\simeq D$.} 
\end{center}
\end{figure}

Thus far we have taken the magnetic field to be completely perpendicular to the DM vector.
Figs.~\ref{fig:fig5} and ~\ref{fig:fig_hz} evidence that even a tiny $h_z$ can modify the phase diagram substantially, owing to the large correlation length of the ordered states near $\Delta=1$.
Let us denote as $l_c$ the value of $l$ at which $|y_C|\simeq 1$.
It follows from the analytical solution of the KT equations\cite{gogolin1998} that  
$l_c=[\pi-\cos^{-1}(-y_\sigma(l_\phi)/y_C(l_\phi))]/\sqrt{y_C(l_\phi)^2-y_\sigma(l_\phi)^2}$ 
(for $|y_C(l_\phi)|>|y_\sigma(l_\phi)|$) 
or $l_c=\cosh^{-1}(y_s(l_\phi)/|y_C(l_\phi)|)/\sqrt{y_\sigma(l_\phi)^2-y_C(l_\phi)^2}$
(for $|y_C(l_\phi)|<|y_\sigma(l_\phi)|$).
We find $l_c>10$ for typical values of $D$ and $h_x$.\cite{l_c}
This means that the correlation length of the antiferromagnetic states is $\xi> a_0 \exp(10)$. 
When $h_z\neq 0$, $t_\theta$ is finite and insofar as $t_\theta\leq \exp(-10) a_0^{-1}$ the flow of $y_C$ stops before it reaches the strong coupling limit. 
Specifically, the critical value for the field is
\begin{equation}
\label{eq:hzc}
\frac{h_{\rm z,c}}{h_x}\simeq\frac{T_0}{D} e^{-l_c}<<1,
\end{equation}
which we plot in Fig.~(\ref{fig:hzc}).
All in all, this figure indicates that the field-induced long-range order (``$N^z$'') is hardly detectable experimentally when $\Delta\lesssim 1$.
For small values of $h_z$, the typical flow diagrams for $h_z\neq 0$ are identical to those of Figs.~\ref{fig:frg1},~\ref{fig:frg2} and~\ref{fig:frg3} except for the fact that now the plots must end at $l=l_\theta$. 
As an example, consider the parameters of Fig.~\ref{fig:frg3} with $h_z=10^{-6} h_x$. 
It follows that $l_\theta\simeq 16$. 
The ground state is LL because $|y_C(16)|<<1$, as shown in Fig.~\ref{fig:frg3}.

\subsection{Abelian bosonization}

The main limitation of the non-Abelian bosonization method employed so far is that it is designed for SU(2) symmetric systems, thus becoming gradually unreliable away from $\Delta_{\rm eff}=1$.
In this section we attempt to overcome this limitation by turning to Abelian bosonization, which is better suited to handle exchange anisotropy.
Nonetheless, the Abelian bosonization approach has shortcomings of its own.
Most notably, there is no obvious way to perform the chiral rotation that proved very helpful in the non-Abelian case.
All in all, this subsection does not presents new results but rather revisits from a different viewpoint the overall features of the phase diagram derived above.

We begin by gauging away the DM interaction as explained in Section I; this renormalizes the anisotropy parameter $\Delta\to\Delta_{\rm eff}=\Delta/\sqrt{1+D^2/J^2}$ and introduces spatial oscillations in the component of the magnetic field that is perpendicular to the DM vector.
Next, by mediation of the Jordan-Wigner transformation we express ${\bf S}_j$ in terms of spinless fermions $\psi_j$;
 Eq.~(\ref{eq:h1}) then turns into  ${\cal H}={\cal H}_0+{\cal H}_{\rm bs}+{\cal V}$, where
\begin{equation}
{\cal H}_0=J\sum_j\frac{1}{2}\left(\psi^\dagger_j\psi_{j+1}+{\rm h.c.}\right)
\end{equation}
is the non-interacting part,
\begin{equation}
{\cal H}_{\rm bs}=J \Delta_{\rm eff}\sum_j\left(\psi^\dagger\psi_j-\frac{1}{2}\right)\left(\psi^\dagger_{j+1}\psi_{j+1}-\frac{1}{2}\right)
\end{equation}
is the interacting part ($\Delta_{\rm eff}$ plays the role of interaction strength), 
\begin{eqnarray}
{\cal V}&=&-h_z\sum_j \left(\psi^\dagger_j\psi_j-\frac{1}{2}\right)\nonumber\\
&-&h_x\sum_j \left(\psi_j^\dagger e^{i\alpha j}e^{i\pi\sum_{k<j}\psi^\dagger_k\psi_k}+{\rm h.c.}\right)
\end{eqnarray}
collects the DM and Zeeman terms and $\alpha=\tan^{-1}(D/J)$.
The low energy properties of this model may be captured by linearizing the dispersion of the spinless fermions around the Fermi points and going to the continuum limit:
\begin{equation}
\psi_j/\sqrt{a}\simeq R e^{i k_F x} + L e^{-i k_F x}  \mbox{    ;    } \psi_{j+1}\simeq \psi(x)+ a \partial_x \psi(x),
\end{equation}
 where $k_F=\pi/2 a+h_z/v$ is the Fermi momentum (half-filling at $h_z=0$).
The right- and left-moving fermion fields are bosonized via
\begin{equation}
R =\frac{1}{\sqrt{2\pi a}}e^{i\sqrt{4\pi}\phi_{R}}\mbox{   ;    } L =\frac{1}{\sqrt{2\pi a}}e^{-i\sqrt{4\pi}\phi_{L}}
\end{equation}
where $\phi_{R,L}$ are chiral bosonic fields obeying
\begin{eqnarray}
&&[\phi_R,\phi_L]=\frac{i}{4}\nonumber\\
&&[\phi_{R(L)}(x),\phi_{R(L)}(y)]=+(-)\frac{i}{4}{\rm sign}(x-y).
\end{eqnarray}
Symmetric and antisymmetric combinations of the chiral fields constitute the dual fields introduced in the previous subsection:
\begin{equation}
\Phi=\phi_{R}+\phi_{L} \mbox{  ;    } \Theta=-\phi_{R}+\phi_{L}.
\end{equation}
Using $\Phi$ and $\Theta$ the bosonized form of Eq.~(\ref{eq:h1}) reads
\begin{eqnarray}
\label{eq:h_ab0}
{\cal H} &=& {\cal H}_0+{\cal H}_{\rm bs}+{\cal V}\nonumber\\
{\cal H}_0 &=& \frac{v}{2}\left[(\partial_x\Theta)^2+(\partial_x\Phi)^2\right]\nonumber\\
{\cal H}_{\rm bs} &=& -\frac{2\pi v}{(2\pi a)^2} G\cos\left(\sqrt{8\pi K}\Phi+\frac{2 h_z\sqrt{K}}{v} x\right)\nonumber\\
{\cal V} &=& \frac{h_x}{\pi a}\cos\left(\sqrt{2\pi K}\Phi+ \frac{h_z\sqrt{K}}{v} x\right)\nonumber\\
~~~~~~~~~~~~~&\times&\cos\left(\sqrt{\frac{2\pi}{K}}\Theta + \frac{\alpha}{a\sqrt{K}} x\right).
\end{eqnarray}
In the derivation of ${\cal V}$ we have used
\begin{eqnarray}
\label{eq:s_ab}
S^+(x)&=&\frac{e^{-i\sqrt{2\pi/K}\Theta}}{\sqrt{2\pi a}}\left[(-1)^{x/a}+\cos(\sqrt{2\pi K}\Phi)\right]\nonumber\\
S^z(x) &=& -\frac{\sqrt{K}}{\pi}\partial_x\Phi+\frac{(-1)^{x/a}}{\pi a}\cos(\sqrt{2\pi K}\Phi),
\end{eqnarray}
neglecting rapidly oscillating terms and absorbing $h_z$ and $D$ through a shift in $\Phi$ and $\Theta$, respectively. 
In addition, we have rescaled the bosonic fields as $\Phi\to\sqrt{K}\Phi$ and $\Theta\to\Theta/\sqrt{K}$, which enables us to write the non-interacting part ${\cal H}_0$ in the canonical form.
$K$ is the Luttinger parameter that differs from unity due to interactions ($\Delta_{\rm eff}\neq 0$).
 $y_\sigma\equiv 2(1-K)$, which played a central role in the preceding subsection, characterizes the interaction strength.
The $\cos(\sqrt{8\pi K}\Phi)$ term originates from Umklapp scattering events such as $R^\dagger(x+a)L(x+a)L^\dagger(x)R(x)$.
When $h_z\neq 0$ the fermionic system is away from half-filling, hence the onset of spatial oscillations. 
Likewise, $D$ induces spatial oscillations in $\cos(\sqrt{8\pi/ K}\Theta)$, which can be understood by carrying out the bosonization in a rotated frame.\cite{suhas2008} 
  
The Umklapp perturbation has zero conformal spin and its scaling dimension is $d= (\sqrt{8\pi K})^2/4\pi=2 K$, which is irrelevant for $K>1$ ($\Delta_{\rm eff}<1$).
In contrast, ${\cal V}$ contains a product of two operators with nonzero conformal spin $s=\sqrt{2\pi K}\sqrt{2\pi/K}/2\pi=1$.
As a consequence, ${\cal V}$ generates new perturbations in the course of the RG flow, which have to be taken into account.
These perturbations have zero conformal spin and may be derived as indicated in Ref.~[\onlinecite{gogolin1998}];
 the outcome is
\begin{eqnarray}
\label{eq:h_ab}
&&{\cal H}=\frac{v}{2}\left[(\partial_x\Theta)^2+(\partial_x\Phi)^2\right]\nonumber\\
&&+\frac{4 v z}{a^2}\cos\left(\sqrt{2\pi K}\Phi+\frac{h_z\sqrt{K}}{v} x\right)\cos\left(\sqrt{\frac{2\pi}{K}}\Theta+\frac{\alpha}{a\sqrt{K}} x\right)\nonumber\\
&&-\frac{2\pi v}{(2\pi a)^2}G\cos\left(\sqrt{8\pi K}\Phi+\frac{2 h_z \sqrt{K}}{v} x\right)\nonumber\\
&&-\frac{2\pi v}{(2\pi a)^2}\tilde{G}\cos\left(\sqrt{\frac{8\pi}{K}}\Theta + \frac{2 \alpha}{a\sqrt{K}} x\right),
\end{eqnarray}
where $z\equiv h_x a/4\pi v$.
The new perturbation generated from ${\cal V}$ is $\cos(\sqrt{8\pi/K}\Theta)$, with scaling dimensions $d=2/K$.
This perturbation is relevant at $\Delta_{\rm eff}<1$, which combined with the fact that $\cos(\sqrt{8\pi K}\Phi)$ is irrelevant suggests that the ground state should be described by a pinned $\Theta$ field.
This guess is naive, partly because when $2(K-1/K)<1$ (namely $K\in(1,1.28)$) ${\cal V}$ must be considered together\cite{gogolin1998} with the spinless perturbations, which complicates the outcome.
Moreover, there are the oscillatory phases that stop the flow of the coupling constants.   
As in the non-Abelian study, we elicit the ground state of ${\cal H}$ from a multiple-stage renormalization group analysis.

In the first stage of RG the characteristic momenta are larger than $h_z/v$ and $\alpha/a$ (hereafter we neglect factors of order one that multiply the oscillatory phases) and we can set $h_z=D=0$ in Eq.~(\ref{eq:h_ab}).
When $h_z=D=0$ Eq.~(\ref{eq:h_ab}) describes two weakly coupled Luttinger liquids; the corresponding flow equations are known\cite{gogolin1998} to be
\begin{eqnarray}
\label{eq:rg_ab}
\frac{d z}{d l}&=&\left[2-\frac{1}{2}\left(K+\frac{1}{K}\right)\right]z\nonumber\\
\frac{d G}{d l}&=& 2(1-K) G+\left(K-\frac{1}{K}\right) z^2\nonumber\\
\frac{d\tilde{G}}{d l}&=& 2\left(1-\frac{1}{K}\right)\tilde{G}+\left(\frac{1}{K}-K\right)z^2\nonumber\\
\frac{d K}{d l}&=&\frac{K}{2}\left(\tilde{G}^2\frac{1}{K}-G^2 K\right).
\end{eqnarray}

When $z=0$ and $K>1$, $\tilde{G}$ flows towards strong coupling and $G$ flows to weak coupling.
If $\tilde{G}$ reaches $\simeq 1 (-1)$, the ground state is described by an ordered $\Theta$ (disordered $\Phi$) such that $\cos(\sqrt{8\pi/K}\Theta)= 1 (-1)$. 
From Eq.~(\ref{eq:s_ab}), this implies that only the staggered component of $S^x$ ($S^y$) acquires a nonzero expectation value.
Consequently, the ground state is ``$N^x$'' (``$N^y$'').
Conversely, when $z=0$ and $K<1$, $G$ flows towards strong coupling and $\tilde{G}$ flows to weak coupling.
If $G$ reaches $\simeq 1$, the ground state is characterized by an ordered $\Phi$ (disordered $\Theta$) such that $\cos(\sqrt{8\pi K}\Phi)=1$.
From Eq.~(\ref{eq:s_ab}), this implies that only the staggered component of $S^z$ acquires a nonzero expectation value.
Consequently, the ground state is ``$N^z$''.
When $z\neq 0$ the aforementioned trends are less clear, and a more careful analysis is required.
We note in passing that for $\alpha=z=0$ Eq.~(\ref{eq:h_ab}) is the XYZ Thirring model, which as $h_z$ increases undergoes a series of phase transitions from a commensurate (gapped) spin density wave (SDW) to an incommensurate (gapless) SDW back to a commensurate SDW through a spin flop.\cite{giamarchi1988,gogolin1998}

Eq.~(\ref{eq:rg_ab}) is most reliable for $\Delta_{\rm eff}\simeq 0$ and $h_x,h_z,D<<J$, because under these conditions the initial values for $z$,$G$, $\tilde{G}$ and $y_\sigma$ are guaranteed to be small.
In effect, $z(0)\propto h_x/J$, $G(0)\propto\Delta_{\rm eff}+ \beta z^2$ and $\tilde{G}(0)\propto z^2$, where $\beta$ is a constant that may be derived perturbatively.\cite{suhas2008}
$K(0)$ can be reliably determined by integrating Eq.~(\ref{eq:rg_ab}) {\em backwards} so that for $h_x=h_z=0$ one reproduces the well-established $LL$ ground state with $K(\infty)^{-1}=1-\cos^{-1}(\Delta_{\rm eff})/\pi$ and $G(\infty)=\tilde{G}(\infty)=z(\infty)=0$.
For weak magnetic fields, one may still use the same $K(0)$ to a good approximation.
When $\Delta_{\rm eff}\simeq 0$, $K$ renormalizes little and hence $K(0)\simeq 1-2\Delta_{\rm eff}/\pi$.

In any event, we are most interested in accessing the strongly interacting  regime $\Delta_{\rm eff}\lesssim 1$ so that we can make contact with the previous subsection.
In particular we wish to find out how robust the field-induced ``$N^z$'' ground state is when the easy-plane anisotropy in enhanced. 
Unfortunately, for $\Delta_{\rm eff}\lesssim 1$ the initial values for the coupling constants are uncertain.
The underlying reason is that $G(0)\simeq 1$, which renders Eq.~(\ref{eq:rg_ab}) invalid.
A more sensible approach is to assume that there has been some prior renormalization group flow (with unknown flow equations), which starting from strong coupling has culminated in a relatively small value of $G$ at some $l=l_0$.
The rationale behind this assumption is that for $\Delta_{\rm eff}<1$ the $\cos(\sqrt{8\pi K}\Phi)$ term is irrelevant.
Thereafter Eq.~(\ref{eq:rg_ab}) determines the flow at $l>l_0$, and we are left to guess the initial conditions of the coupling constants at $l=l_0$.
We take $z(l_0)\propto h_x/J$ and  $\tilde{G}(l_0)\propto z(l_0)^2$, with proportionality constants of order unity.
On the other hand, we {\em choose} the value for $K(l_0)$ by hand; this is tantamount to selecting an intermediate energy scale for $l=l_0$, which corresponds to a lengthscale $a_0$ that is larger than the lattice constant $a$.
Finally, we integrate Eq.~(\ref{eq:rg_ab}) backwards to determine $G(l_0)$ such that the ground state in absence of fields will reproduce $G\to 0$ and $K\to K_{\rm inf}$.
Since $h_z,h_x<<J$, presumably the value of $G(l_0)$ will be nearly independent of the magnetic field.  
Overall, our choice of $K(l_0)$ is engineered in a way that Eq.~(\ref{eq:rg_ab}) will reproduce the known ground states of a variety of limiting cases, without having to tune the initial values for the coupling constants. 
These limiting cases are

(i) $\Delta_{\rm eff}<1$, $h_x=h_z=0$, any $D$.

In this case $z(l)=\tilde{G}(l)=0$ and we are left with flow equations for $G$ and $K$.
We find that the ground state is $LL$, in agreement with Bethe ansatz calculations.

(ii) $\Delta_{\rm eff}<1$, $h_z=D=0$ and $h_x\neq 0$. 

In this case $\tilde{G}$ flows to strong coupling: $\tilde{G}\to -\infty$, $G\to 0$, $K\to\infty$.
This corresponds to the ``$N^y$'' ground state, which is the expected answer as discussed in the previous subsection.

(iii) $\Delta_{\rm eff}<1$, $h_x=0$, $h_z\neq 0$, any $D$.

Here $z(l)=0$. 
In the first stage of RG ($l<l_1={\rm min}(\log(v/a_0 h_z),\log(a/ a_0\alpha)$) $\tilde{G}$ is the relevant perturbation.
However, because $\tilde{G}(l_0)\propto z(l_0)^2$ and $z(l_0)\propto h_x$, we have $\tilde{G}(l_0)=0$ and $dG(l_0)/dl=0$.
Therefore $\tilde{G}(l)=0$ and there is no possibility for a ``$N^y$'' ground state.
The effective RG equations for the first stage are thus
\begin{eqnarray}
\frac{d G}{d l}=2(1-K) G\nonumber\\
\frac{d K}{d l}=-\frac{1}{2} K^2 G^2.
\end{eqnarray}
Since $K(l_0)>1$, $G$ decreases (it is irrelevant at $\Delta_{\rm eff}<1$); so does $K$, but more slowly than $G$.
Therefore $G$ cannot reach the strong coupling limit either and moreover its flow stops at $l=l_1$.
In sum, the ground state is $LL$, which agrees with Bethe ansatz results.

(iv) $\Delta_{\rm eff}=1$, $D\neq 0$, $h_x\neq 0$ and $h_z=0$. 

$\Delta_{\rm eff}=1$ is the situation for which the non-Abelian bosonization discussed above is reliable.
In this case $z$ and $\tilde{G}$ stop flowing at $l=\log(a/a_0 \alpha)$, beyond which we can set $z=\tilde{G}=0$ in Eq.~(\ref{eq:rg_ab}) and keep integrating the flow equations for $K$ and $G$.
We obtain $G\to\infty$ and $K\to -\infty$ regardless of the $h_x/D$ ratio. 
This corresponds to the ``$N^z$'' ground state and is in agreement with the results derived in the previous subsection.

\begin{figure}
\begin{center}
\scalebox{0.3}{\includegraphics{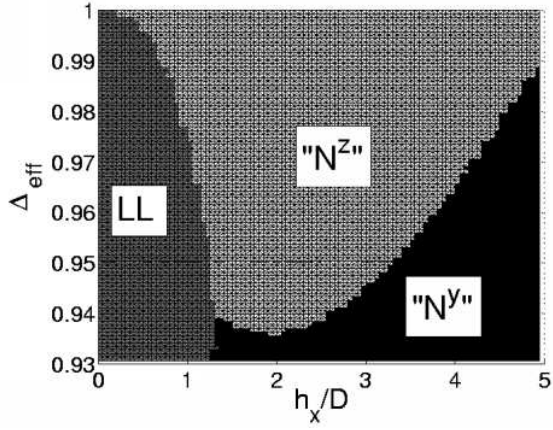}}
\caption{Influence of exchange anisotropy on the phase diagram of Eq.~(\ref{eq:h1}); derived using Abelian bosonization. $D=0.01 J$, $h_z=0$. The initial energy scale for the RG flow was chosen via $K(l_0)=1.1 K(\infty)$. For $\Delta_{\rm eff}\lesssim 0.97$ only ``$N^y$'' and $LL$ phases can be found. This plot agrees roughly with Fig.~\ref{fig:fig2}, which was derived using non-Abelian bosonization. Disagreements between the figures are most noticeable on the shape of phase boundaries.} 
\label{fig:ab1}                                                                                                                                                               
\end{center}
\end{figure}

\begin{figure}
\begin{center}
\scalebox{0.3}{\includegraphics{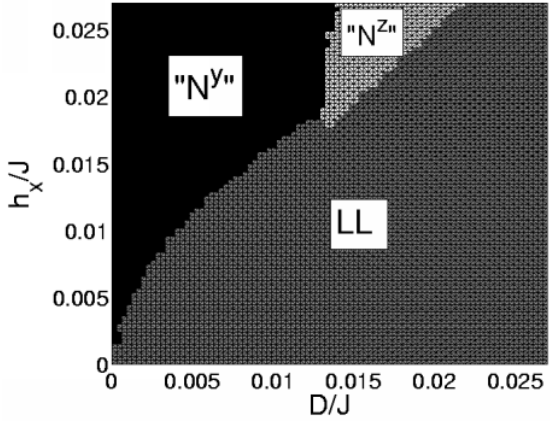}}
\caption{Phase diagram for $h_z=0$, $\Delta=0.93$; derived using Abelian bosonization. The initial energy scale for the RG flow was chosen via $K(l_0)=1.1 K(\infty)$. For such $\Delta$, ``$N^z$'' is absent for $D\lesssim 0.015 J$ (see also Fig.~\ref{fig:ab1}); it reappears at larger $D$ values. This demonstrates that $\Delta_c$ gets smaller as $D$ and $h_x$ increase.}
\label{fig:ab2}                                                                                                                                                               
\end{center}
\end{figure}

\begin{figure}
\begin{center}
\scalebox{0.3}{\includegraphics{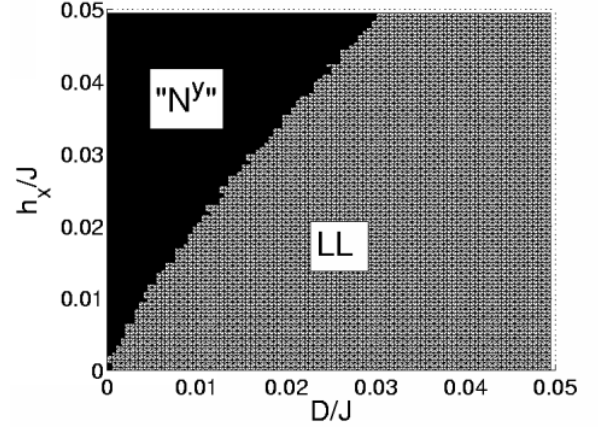}}
\caption{Phase diagram in the neighborhood of the non-interacting limit ($\Delta=0.1$), where the Abelian bosonization scheme utilized is most reliable. ``$N^y$'' is defined via $\tilde{G}(l_1)>0.15$; other choices would lead to a quantitative change in the slope of the phase boundary between ``$N^y$'' and $LL$. Regardless, there is no remnants of the ``$N^z$'' phase.}
\label{fig:ab3}                                                                                                                                                               
\end{center}
\end{figure}

Next we compute the more general phase diagram numerically.
We begin by taking $h_z=0$.
Fig.~\ref{fig:ab1} displays the three phases that compete with each other when $\Delta_{\rm eff}<1$, and agrees roughly with Fig.~\ref{fig:fig2} derived from non-Abelian bosonization.
Below $\Delta_{\rm eff}\leq\Delta_c$, ``$N^z$'' is no longer the ground state; instead, ``$N^y$'' and $LL$ are stabilized depending on the $h_x/D$ value.
When $D=0$ and $h_x\neq 0$, the ground state is invariably ``$N^y$''.
Likewise, when $D\neq 0$ and $h_x=0$, the ground state is invariably $LL$.
However, ``$N^z$'' can be the ground state at $D\neq 0\neq h_x$ and $\Delta_{\rm eff}<1$, even though $\cos(\sqrt{8\pi K}\Phi)$ is irrelevant, because of the frustrating influence that $D$ and $h_x$ have on each other.
Similarly, a $LL$ ground state may emerge at $h_x\neq 0$ due to the DM interaction, which may cut off the flow of $\tilde{G}$ before it reaches the strong coupling regime.
We use $\tilde{G}(l_1)\gtrsim 0.15$ as the criterion that defines ``$N^y$'' (note that $\tilde{G}(l_0)<<1$). 
This is a somewhat arbitrary choice that endows the phase boundaries with uncertainty; nevertheless it is motivated as an attempt to reproduce the phase diagrams of the previous subsection. 

Unlike in the previous subsection, in the present context the ``$N^y$'' phase displays no trace of dimerization.
In effect, here dimerization is associated with the ordering of $\Phi$ such that $\cos(\sqrt{8\pi K}\Phi)=-1$; 
yet in the ``$N^y$'' phase $\Phi$ is disordered because $\Theta$ is pinned.
The fact that non-Abelian bosonization is able capture the coexistence of antiferromagnetism and dimerization is more a merit of the chiral rotation than an intrinsic  flaw of the Abelian bosonization.

Fig.~\ref{fig:ab2} sheds light on the parameter space for which ``$N^z$'' constitutes the ground state.
For given $\Delta_{\rm eff}<1$ and $h_x/D$, $N_z$ is more robust at larger $D$ (or $h_x$).
In other words, $\Delta_c$ decreases as $D$ and $h_x$ increase and their ratio is kept of order one; this is in agreement with the results derived in the previous subsection.
The reason behind this trend is that $G$ and $\tilde{G}$ compete which each other, the latter being the most relevant perturbation at $\Delta_{\rm eff}\lesssim 1$.
The larger $D$ is the less chance $\tilde{G}$ has to reach strong coupling, thereby increasing the likelihood for $G$ to prevail.

When $h_z\neq 0$, the flow of $G$ is interrupted.
In qualitative agreement with Fig.~\ref{fig:fig5}, we find (not shown) that the parameter range for ``$N^z$'' narrows significantly even for a small value of $h_z$.
However, we remark that in the present formalism $h_z$ hampers ``$N^z$'' but not ``$N^y$'' because $h_z$ does not induce oscillations in the factor that multiplies $\tilde{G}$. 
This is in qualitative disagreement with the outcome of non-Abelian bosonization, where $h_z\neq 0$ reduces the likelihood of {\em both} ``$N^y$'' and ``$N^z$'' (because both phases are linked to the {\em same} coupling constant $y_C$, cf. Sec. IVa).
The classical study of Section III rules in favor of the non-Abelian result by anticipating an increase of the critical field for the spin-flop transition when $h_z\neq 0$.

As mentioned above, the main results of this subsection suffer from uncertainties in the initial conditions for the coupling constant.
This problem is remedied in the weakly interacting limit ($\Delta_{\rm eff}\simeq 0$) for which the initial values of the coupling constants are small and known.
Fig.~\ref{fig:ab3} demonstrates that for large easy-plane anisotropy ($\Delta=0.1$) only ``$N^y$'' and $LL$ can be the ground states.
This, in conjunction with the non-Abelian study of the SU(2)-symmetric point, ratifies that there exists a critical value of $\Delta$ below which ``$N^z$'' disappears.
While our calculations indicate that $\Delta_c$ is close to one, numerical density-matrix RG studies might be desirable to ascertain its precise value, as well as to corroborate the coexistence of dimerization and antiferromagnetism in the ``$N^y$'' phase.

\section{Application to superconductivity}

Thus far we have discussed the ground states of Eq.~(\ref{eq:h1}) in the context of one-dimensional quantum spin chains.
However, there appear to be few experimental studies on one-dimensional antiferromagnets with uniform DM interactions.
In this section we demonstrate that Eq.~(\ref{eq:h1}) also models superconducting nanostructures that might be realizable in experiments.

First, consider a one-dimensional array of Josephson junctions\cite{jj} separated from one another by a distance $a$.
Its Hamiltonian is
\begin{equation}
\label{eq:h_jj}
{\cal H}_{\rm JJ}=\frac{1}{2}\sum_{i,j} n_i C_{i,j}^{-1} n_j -E_J \sum_j \cos(\chi_j-\chi_{j+1}),
\end{equation}
where $n_i$ denotes the number of Cooper pairs in the $i$-th superconducting island, $\chi_i$ is the U(1) superconducting angle for the $i$-th island (canonically conjugate to $n_i$), $C_{i,j}$ is the capacitance matrix that models the repulsive Coulomb interactions between the Cooper pairs, and $E_J$ is the Josephson coupling energy.
For conventional Josephson junctions $E_J>0$, while for $\pi$-junctions\cite{frolov2007} $E_J<0$.
We neglect dissipative processes (e.g. quasiparticle tunneling), which are relatively unimportant at low temperatures.
We are interested in small superconducting grains\cite{watanabe2002} where the onsite Coulomb interaction is strong, i.e. $e^2 C_{i,i}^{-1}>>E_J, T$.
Accordingly the superconducting islands are in the Coulomb blockade regime and the large electrostatic energy cost for changing the number of Cooper pairs on each island drives the JJ array to an insulating regime.
We consider the particular case in which $n_i$ can acquire only two possible values; the practical implementation of this scenario may require tuning the chemical potential of the Cooper pairs via a gate voltage.
The two possible values of charge define a pseudospin degree of freedom for each island, which enables the mapping of Eq.~(\ref{eq:h_jj}) into a pseudospin Hamiltonian:\cite{bruder1993,glazman1997} 
\begin{eqnarray}
{\cal H}_{\rm JJ}&=&-E_J\sum_j\left(S^x_j S^x_{j+1}+S^y_j S^y_{j+1}\right)+E_c \sum_j  S^z_j S^z_{j+1}\nonumber\\
&-&h_z\sum_j S^z_j.
\end{eqnarray}
$S^z_j=n_j$ is the number operator for Cooper pairs, and $E_c S^z_{j+1} S^z_{j+1}$ describes the intergrain Coulomb repulsion ($E_c>0$).
We have assumed screened Coulomb interactions,\cite{bradley1984} whereby $C_{i,i}^{-1}>>C_{i,i+1}^{-1}$ and $C_{i,i+n}^{-1}=0$ for $n\geq 2$.
This requires that the self-capacitance of the superconducting island $C_0$ be larger than the junction capacitance $C$.
$C_0$ and $C$ are defined via $C_{i,j}\simeq (C_0+2 C) \delta_{i,j}-C (\delta_{i,j+1}+\delta_{i,j-1})$. 
$S^x_j$ and $S^y_j$ are the real and imaginary part of the superconducting pair operator for the $j$-th grain, and $ E_J(S^x_j S^x_{j+1}+S^y_j S^y_{j+1})$ is associated with the tunneling of Cooper pairs between neighboring grains.
$h_z$ is a pseudospin magnetic field that describes the deviation of the chemical potential from the middle point between the electrostatic energies of the two charge states.
Unlike in quantum antiferromagnets, in JJ arrays $E_J/E_c$ may be tuned {\em in situ}.\cite{sharifi1993}

\begin{figure}
\begin{center}
\scalebox{0.35}{\includegraphics{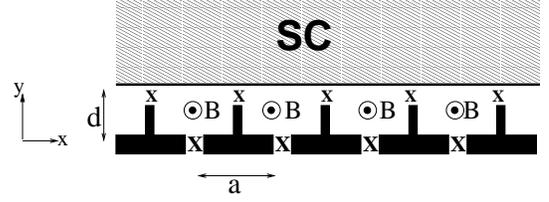}}
\caption{Superconducting analogue of Eq.~(\ref{eq:h1}): a one-dimensional array of small superconducting islands (in black) separated by Josephson junctions (crosses), located in close proximity to a bulk superconductor (shaded area). 
The applied magnetic field (perpendicular to the page) leads to an effective DM interaction in the JJ array.
The Josephson coupling between the bulk superconductor and the array plays the role of a XY magnetic field in pseudospin space; gate voltages are Z magnetic fields in pseudospin space. The pseudospin anisotropy is defined by the disparity between the Josephson coupling and the capacitive energy of the junctions.}
\label{fig:JJ_array}                                                                                                                                                             \end{center}
\end{figure}

Second, let us place a large superconductor parallel to the array of junctions (see Fig.~\ref{fig:JJ_array}), separated by a distance $d$.
If the material placed between the JJ array and the bulk superconductor is a normal metal, a Josephson coupling will ensue as long as $d\lesssim\xi_N$, where $\xi_N$ is the coherence length of the normal metal.
In magnetic language, the influence of the large superconductor is equivalent to that of an external magnetic field oriented in the $xy$ plane:
\begin{equation}
\label{eq:xxz_h}
{\cal H}={\cal H}_{\rm JJ}-h_x\sum_j S^x_j,
\end{equation}
where our gauge choice is determined by $h_y\equiv 0$.
$h_x$ is proportional to the mean field order parameter of the large superconductor. 
Unlike in the JJ array, we neglect phase fluctuations in the bulk superconductor.
Eq.~(\ref{eq:xxz_h}) is a XXZ model with a {\em uniform} pseudo-magnetic field.

Third, we add the ingredient which will result in a Dzyaloshinskii-Moriya interaction in pseudospin space.\cite{gingras1992}
Let us apply a uniform magnetic field ${\bf B}=B\hat z$ (see Fig.~\ref{fig:JJ_array}).
The vector potential ${\bf A}$ associated with the magnetic field twists the superconducting angle, so that  
\begin{eqnarray}
\label{eq:h000}
{\cal H}&=&-\frac{E_J}{2}\sum_j \left(e^{i \theta} S^+_j S^-_{j+1}+{\rm h.c.}\right)+E_c\sum_j S^z_j S^z_{j+1}\nonumber\\
&-&h_x\sum_j \left(e^{i\tilde{\theta} j} S^+_j+{\rm h.c.}\right)-h_z\sum_j S^z_j.
\end{eqnarray}
where $\theta=(2\pi/\Phi_0)\int_{j}^{j+1}{\bf A}(y=0)\cdot d{\bf x}$ and 
$\tilde{\theta}=(2\pi/\Phi_0)\int_{j}^{j+1}{\bf A}(y=d)\cdot d{\bf x}$. 
$\Phi_0=h/2e=2\times10^{-15} {\rm Wb}$ is the flux quantum and we have taken $y\equiv 0$ at the location of the JJ array.
Eq.~(\ref{eq:h000}) assumes that in spite of the vector potential the superconducting phase is spatially homogeneous {\em within each island}.
This is a reasonable approximation insofar as the magnetic flux threading the island is small.
For convenience we use ${\bf A}= -B y \hat x$, which results in 
\begin{eqnarray}
\label{eq:h_sc2}
{\cal H}&=&-\frac{E_J}{2}\sum_j \left(S^+_j S^-_{j+1}+{\rm h.c.}\right)+E_c\sum_j S^z_j S^z_{j+1}\nonumber\\
&-& h_x\sum_j \left(e^{i\tilde{\alpha} j} S^+_j+{\rm h.c.}\right)-h_z\sum_j S^z_j,
\end{eqnarray}
with 
\begin{equation}
\tilde{\alpha}=-2\pi\frac{\Phi_B}{\Phi_0} \mbox{   ;    }\Phi_B\equiv B d a.
\end{equation}
$\Phi_B$ is the magnetic flux penetrating a rectangle formed by a Josephson junction, perpendicular lines to the bulk superconductor and the edge of the bulk superconductor. 
Our choice of the vector potential corresponded to a spatially uniform magnetic field; nevertheless, Eq.~(\ref{eq:h_sc2}) is valid more generally. 
If $E_J<0$ ($\pi$-junctions), Eq.~(\ref{eq:h_sc2}) completes the mapping into Eq.~(\ref{eq:h2b}).
In contrast, if $E_J>0$ (conventional junctions) we need to make an additional pseudospin rotation by an angle $\pi$ for every other site: $S^+_j\to \exp(i\pi j) S^+_j$.
The resulting correspondence between the antiferromagnetic and the superconducting models can be summarized as follows:
$\begin{array}{ccc}
\\
\mbox{Antiferromagnet} & \mbox{   Conventional JJ Array} &\mbox{   $\pi$ - JJ Array}\\
\hline\\
\sqrt{J^2+D^2} & E_J & -E_J\\
J\Delta & E_c & E_c\\
\alpha=\tan^{-1}(D/J) & \tilde{\alpha}+\pi & \tilde{\alpha}\\
h_x, h_z   &   h_x, h_z  & h_x, h_z\\
\mbox{ } & \mbox{} & \mbox{}
\end{array}$
\\
The phase diagrams calculated in the previous sections are valid for $D,h_x,h_z<<J$.
Therefore, those results may be transferred directly to the case of conventional junctions only when $\Phi_B\simeq\Phi_0/2$ and $h_x,h_z<<E_J$.
On the other hand, for $\pi$-junctions our RG analysis has access to $\Phi_B\simeq 0$ and $h_x,h_z<<E_J$.

\begin{figure}
\begin{center}
\scalebox{0.35}{\includegraphics{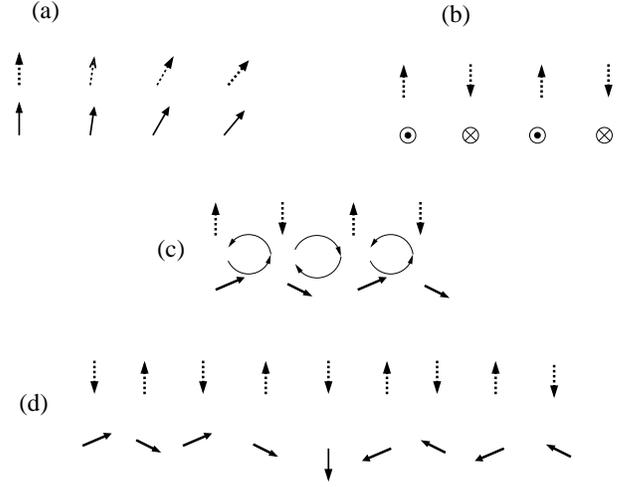}}
\caption{Analogues of the magnetic phases in {\em conventional} Josephson junction arrays -- classical representation for $h_z=0$. Dashed arrows represent the direction of the order parameter ($\tilde\alpha j$) in the bulk superconductor. The gauge choice is determined  by Eq.~(\ref{eq:h_sc2}).  Solid arrows portray the order parameter of the superconducting islands in the one-dimensional JJ array. The vector potential twists the direction of the pseudospins in the XY plane.
(a) Ferromagnetic phase. This is the classical ground state for small magnetic fields ($\Phi_B\to 0$), a regime in which the field theoretical results of the present paper do not apply. The spatial gradient of the order parameter implies Meissner currents flowing at the edge of the bulk superconductor.
(b) ``$N^z$'' phase at $\Phi_B\simeq\Phi_0/2$ (for simplicity we have plotted the order parameter of the bulk superconductor as though $\tilde\alpha=\pi$). This charge density wave state is dominant at $E_c>E_J$ and may also arise at $E_c\lesssim E_J$ provided that $|\Phi_B/\Phi_0-1/2|\simeq h_x/E_J$ (although the experimental detection in the latter regime is highly unlikely). The pseudospins of the array point along Z with alternating sign.
(c) ``$N^y$'' phase at $\Phi_B\simeq\Phi_0/2$ (for simplicity we have plotted the order parameter of the bulk superconductor as though $\tilde\alpha=\pi$). There are circulating supercurrents of alternating chirality and oscillating magnitude across the interface between the bulk superconductor and the 1D array. 
(d) $LL$ phase at $\Phi_B\simeq\Phi_0/2$ (for simplicity we have plotted the order parameter of the bulk superconductor as though $\tilde\alpha=\pi$). When $\Phi_B-\Phi_0/2=0$ the array ignores the twist in the order parameter of the bulk superconductor and adopts a ferromagnetic configuration (not shown) along an arbitrary direction in the XY plane, much as though the bulk superconductor did not exist.
When $0\neq|\Phi_B-\Phi_0/2|<h_x/E_J$ the classical configuration is a soliton lattice (only one soliton is shown in the figure).
} 
\label{fig:JJ_phases}
\end{center}
\end{figure}

Fig.~\ref{fig:JJ_phases} illustrates the physical meaning of the magnetic phase diagram in the present context.
``$N^z$'' corresponds to an insulating charge density wave phase, where the number of Cooper pairs oscillates from one island to another.
``$N^y$'' is a vortex phase\cite{teitel1983} where there are circulating currents with alternating chirality flowing between the JJ array and the bulk superconductor.
In addition, ``$N^y$'' contains some dimerization:
\begin{eqnarray}
\label{eq:dim}
\langle S^+_j S^-_{j+1}\rangle &=& \alpha+(-1)^j \beta\nonumber\\
\langle S^z_j S^z_{j+1}\rangle &=& \gamma+(-1)^j \eta,
\end{eqnarray}
where $\alpha,\beta,\gamma,\eta$ are constants.
The first line of Eq.~(\ref{eq:dim}) implies that the effective Josephson coupling between sites $ 2 j$ and $2 j-1$ is larger than that between sites $2 j$ and $2 j+1$.
In other words, the magnitude of the circulating currents oscillates from one ``plaquette'' to another and  is larger for one chirality than for the opposite chirality.
The second line of Eq.~(\ref{eq:dim}) means that the magnitude of the effective junction capacitance oscillates from one plaquette to another.
In other words, if island $2 j$ has zero Cooper pairs then site $2 j-1$ is more likely to have one Cooper pair than site $2 j+1$.
Finally, $LL$ is the gapless ground state in which the superconducting angles of the junctions form a soliton lattice.
Throughout the foregoing discussion we have generally ignored the back action of the 1D array on the bulk superconductor, and in particular we have neglected the magnetic fields generated by the alternating currents in the ``$N^y$'' phase.
This approximation is safest when the effective London penetration depth is larger than the size of the array.\cite{tinkham1996}  

Let us discuss the phase diagram of a conventional array with $E_c<E_J$, which in the magnetic problem corresponds to $\Delta_{\rm eff}<1$.
When $\Phi_B\simeq 0$, the order parameter of the JJ array is aligned ferromagnetically with that of the bulk superconductor.
Although this result does not follow from the field theoretical calculations of the present paper, its counterpart in quantum spin chains is well-established.
In effect, for $\tilde{\alpha}=0$ and $h_z=0$ the superconducting model becomes equivalent to a spin 1/2 antiferromagnet with {\em staggered} DM interaction in a transverse magnetic field, which was first studied in Ref.~[\onlinecite{affleck1999}].
The ground state in this case was found to be antiferromagnetic with the Neel vector aligned with the external field.
In pseudospin language this translates into the aforementioned ferromagnetic ground state.

As the magnetic field increases the pseudospins of the JJ array are increasingly twisted, with the concomitant loss of exchange energy.
For $\Phi_B\simeq \Phi_0/2$, it is no longer optimal to have a ferromagnetic alignment between the array and the bulk superconducting angles.
Instead, the ground state is ``$N^y$'' (if $|\Phi_B/\Phi_0-1/2| < h_x/E_J$) or  $LL$ (if $|\Phi_B/\Phi_0-1/2| > h_x/E_J$).
Furthermore, if $E_c/E_J\in(\Delta_c,1)$, ``$N^z$'' emerges as the ground state at $|\Phi_B/\Phi_0-1/2|\simeq h_x/E_J$.
Finding ``$N^z$'' at $E_c<E_J$ is counterintuitive because it means that the proximity coupling from the bulk superconductor drives the array into an {\em insulating} ground state.
By tuning $h_x$ or the applied (real) magnetic field, one may induce transitions between the three phases.
In particular, if $E_c/E_J<\Delta_c$ there is a commensurate-incommensurate transition between $LL$ and ``$N^y$''.
The critical value of the proximity coupling for this transition is 
\begin{equation}
h_{x,c}\simeq \frac{\pi}{2}(\tilde\alpha+\pi) E_J.
\end{equation}
This critical field changes in presence of a gate voltage ($h_z$) as indicated by Eq.~(\ref{eq:hc2}).
 
The experimental detection of the aforementioned phases requires SQUID measurements,\cite{frolov2007} which would target the circulating currents of ``$N^y$'' ground states, as well as measurements of the critical current of the array,\cite{watanabe2002} which would be exponentially suppressed with the length of the array in the ``$N^z$'' ground state\cite{glazman1997} but not in the $LL$ phase.
We briefly comment on a number of additional experimental requirements:

(i) The Josephson coupling between the islands in the array must be stronger than the coupling between the islands and the bulk superconductor, because our field theoretical results apply for $h_x<<E_J$. Moreover, our model applies for short-ranged Coulomb interactions in the array, i.e. for $C<C_0$.
  
(ii) The temperature of the system must be smaller than the gaps in the ``$N^y$'' and ``$N^z$'' ground states: $T<<h_x<<E_J,E_c$.
For typical values of the Josephson coupling ($E_J\lesssim 1 K$ for low-temperature superconductors) this requirement is most pressing at $E_c\simeq E_J$, where the energy gap associated with ``$N^z$'' is only $\simeq 0.077 E_J \exp(-10)\simeq 5 \mu K$ (recall Section IVa).
As $E_c$ and $E_J$ are made dissimilar the gaps may increase to $O(h_x)\lesssim 10 mK$ .

(iii) The array must be long enough so as to reduce the quantum tunneling between degenerate ground states. 
In effect, ``$N^y$'' and ``$N^z$'' each break a $Z_2$ symmetry: there are two degenerate ``$N^z$'' or ``$N^y$'' phases that differ from each other only by a translation of a lattice constant.

(iv) The superconducting grains must be small enough to justify the pseudospin 1/2 approximation. 
At the same time the area of a plaquette ($a\times d$) should be large enough to enclose a flux $\Phi_B=\Phi_0/2$ using magnetic fields that are smaller than the critical field of the superconducting islands (note that this concern does not apply to arrays of $\pi$-junctions, for which we require $\Phi_B\simeq 0$).
The ``inverted-T'' shape of the islands depicted in Fig.~\ref{fig:JJ_array} could help satisfy both conditions.
For aluminum, $a\times d\gtrsim 0.1 \mu m^2$ would ensure that the applied field remains below the critical field.\cite{delsing1994}



\section{Summary and Conclusions}

We have evaluated the zero-temperature phase diagram of an antiferromagnetic spin 1/2 chain in presence of uniform Dzyaloshinskii-Moriya interactions,  symmetric exchange anisotropy and arbitrarily oriented magnetic fields.
We have used non-Abelian as well as Abelian bosonization, and have generally found qualitative agreement between the two schemes.
When the two diverge, the former approach proves to be more reliable.
Our calculations predict the emergence of three competing phases for spin chains with easy-plane anisotropy.
One of them (phase (i)) is an antiferromagnet with its Neel vector along the direction of the DM vector.
This phase was introduced in previous work,\cite{suhas2008} whose scope was limited by the assumption of isotropic symmetric exchange.  
Our results indicate that phase (i) is unstable under weak-to-moderate easy-plane anisotropy: we have estimated the critical value of the anisotropy beyond which it disappears.
This value is sensitive to the magnitude of the DM interaction, as well as to the magnitude and direction of the applied magnetic field.
The two new ground states that occur as a consequence of symmmetric easy-plane exchange anisotropy are (ii) a dimerized antiferromagnet with Neel vector perpendicular to both the DM vector and the magnetic field, (iii) a gapless Luttinger liquid, whose classical counterpart is a soliton lattice.
Phase (ii) arises when the DM interaction is weak compared to the magnetic field component transverse to the DM vector; phase (iii) ensues in the opposite regime.
Phase (i) may then be understood as an outcome of the frustration between competing phases (ii) and (iii); indeed it is most likely to emerge when the DM interaction is neither large nor small compared to the transverse magnetic field component. 
It would be interesting to verify and refine these predictions using the numerical density-matrix renormalization group method.

Motivated in part by the scarcity of experiments on one-dimensional antiferromagnets with uniform Dzyaloshinskii-Moriya interaction, we have searched for alternative systems where our calculations may be experimentally tested.
Thus we have mapped the original magnetic problem into a mathematically equivalent superconducting problem involving a one-dimensional array of Josephson junctions (either conventional or $\pi$-type junctions) in close proximity to a bulk superconductor. 
An applied perpendicular magnetic field plays the role of a uniform DM interaction.
We have discussed the physical meaning of the magnetic phases in the superconducting context, including that of dimerization.
The exquisite tunability of Josephson junction parameters in one-dimensional arrays may provide an interesting avenue to probe and replicate the influence of DM interactions and magnetic fields in one-dimensional quantum antiferromagnets with symmetric exchange anisotropy.

\acknowledgements
We thank M. Franz, W. Hardy and H. Karimi for helpful discussions.
This research has been supported by NSERC and CIfAR.
I.G. is a CIfAR Junior Fellow.

\appendix
\begin{widetext}
\section{Influence of $h_z$ on the Classical Commensurate-Incommensurate Transition}

The objective of this Appendix is to determine how the classical soliton lattice (Sec. III) is modified when a magnetic field is applied along the direction of the DM vector.
Let us parametrize the classical spin at site $j$ as ${\bf S}_j=S(\sin\theta_j \cos\tilde{\phi}_j, \sin\theta_j \sin\tilde{\phi}_j, \cos\theta_j)$.
Then Eq.~(\ref{eq:h1}) can be rewritten as
\begin{equation}
{\cal H} =-\tilde{J} S^2 \sum_j\left[ \sin\theta_j \sin\theta_{j+1} \cos(\phi_{j+1}-\phi_j-\alpha)-\Delta_{\rm eff} \cos\theta_j \cos\theta_{j+1}\right] - h_x S \sum_j (-1)^j \sin\theta_j \cos\phi_j - h_z S \sum_j \cos\theta_j,
\end{equation}
where $\phi_j=\tilde\phi_j-\pi j$, $\tilde{J}=J \sqrt{1+D^2/J^2}$ and $\Delta_{\rm eff}=\Delta/\sqrt{1+D^2/J^2}$.
Taking advantage of the fact that $\phi_{j+1}-\phi_j-\alpha$ is small for each $j$, we write
\begin{eqnarray}
{\cal H} &=&-\tilde{J} S^2 \sum_j\left[ \sin\theta_j \sin\theta_{j+1} \left(1-\frac{1}{2}(\phi_{j+1}-\phi_j-\alpha)^2\right)-\Delta_{\rm eff} \cos\theta_j \cos\theta_{j+1}\right]\nonumber\\
\
&&- h_x S \sum_j (-1)^j \sin\theta_j \cos\phi_j - h_z S \sum_j \cos\theta_j
\end{eqnarray}
Let us define
\begin{eqnarray}
\phi_j &=& a(j) + (-1)^j b(j)\nonumber\\
\theta_j &=& c(j) + (-1)^j d(j),
\end{eqnarray}
where $a,b,c,d$ are functions that vary slowly along the soliton.

Keeping only the non-alternating terms and making the continuum approximation we arrive at the following expression for the Hamiltonian density ${\mathbb h}$ (${\cal H}\equiv\int d x {\mathbb h}$):
\begin{eqnarray}
{\mathbb h}&=&-\tilde{J} S^2\left\{(1-\beta^2-d^2)\left[1-\frac{1}{2}\left(\frac{d a}{d x}-\alpha\right)^2-2 b^2\right]-\Delta_{\rm eff}(\beta^2-d^2)\right\}\nonumber\\
&&-h_x S\left[\beta d\left(1-\frac{b^2}{2}\right)\cos a - \left(1-\frac{d^2}{2}-\frac{\beta^2}{2}\right) b \sin a\right]\nonumber\\
&&-h_z S \beta\left(1-\frac{d^2}{2}\right)
\end{eqnarray}
where $\beta=\cos c$.
In addition we have used $\sin b\simeq b$, $\cos b\simeq 1-b^2/2$, $\sin d\simeq d$, $\sin c\simeq 1-\beta^2/2$ and so on.

Now we determine the optimal value for the functions $a,b,c,d$.
The value of $b$ that minimizes ${\mathbb h}$ is
\begin{eqnarray}
b &=& -\frac{h_x S \sin a}{4 \tilde{J} S (1-\beta^2-d^2)+h_x \beta d \cos a} \left(1-\frac{d^2}{2}-\frac{\beta^2}{2}\right)\nonumber\\
&\simeq & - \frac{h_x \sin a}{4 \tilde{J} S} \left(1+\frac{\beta^2}{2}+\frac{d^2}{2}-\frac{h_x \beta d \cos a}{4 \tilde{J} S}\right)
\end{eqnarray}
Substituting this expression back in the Hamiltonian we obtain

\begin{equation}
{\mathbb h} \simeq  -\tilde{J} S^2\left\{(1-\beta^2-d^2)\left[1-\frac{1}{2}\left(\frac{d a}{d x}-\alpha\right)^2\right]-\Delta_{\rm eff}(\beta^2-d^2)\right\}-h_x S \beta d \cos a - h_z S \beta \left(1-\frac{d^2}{2}\right)-\frac{h_x^2 \sin^2 a}{8 \tilde{J}}
\end{equation}

Next we optimize $d$.
For $\Delta_{\rm eff}<1$, ${\mathbb h}$ is minimized for $d$ given by
\begin{equation}
d\simeq\frac{h_x \beta \cos a}{2 \tilde{J} S (1-\Delta_{\rm eff})}\left[1+\frac{1}{2(1-\Delta_{\rm eff})}\left(\frac{d a}{d x}-\alpha\right)^2-\frac{ h_z \beta}{2 \tilde{J} S (1-\Delta_{\rm eff})}\right],
\end{equation}
which indicates that the staggered component of $\theta$ is nonzero only near the core of the soliton (far from the core $\cos a\simeq 0$ as shown in Ref.~[\onlinecite{soliton}]).
For simplicity in the above approximation we assumed that $|\Delta_{\rm eff}-1| >> h_z^2/\tilde{J}^2 S^2, (d a/d x-\alpha)^2$.
Substituting the expression for $d$ in the Hamiltonian we get

\begin{equation}
\label{eq:h2}
{\mathbb h}\simeq -\tilde{J} S^2 \left\{(1-\beta^2)\left[1-\frac{1}{2}\left(\frac{d a}{d x}-\alpha\right)^2\right]-\Delta_{\rm eff}\beta^2\right\}-h_z S \beta-\frac{h_x^2 \sin^2 a}{8 \tilde{J}}-\frac{h_x^2 \beta^2 \cos^2 a}{4 \tilde{J} (1-\tilde{\Delta})}.
\end{equation}

Next we optimize $\beta$.
For $\Delta_{\rm eff}<1$, ${\mathbb h}$ is minimized for $\beta$ given by
\begin{equation}
\beta=\frac{h_z}{2 \tilde{J} S (1+\Delta_{\rm eff})}\left[1+\frac{1}{2(1+\Delta_{\rm eff})}\left(\frac{d a}{d x}-\alpha\right)^2+\frac{h_x^2 \cos^2 a}{4 \tilde{J}^2 S^2 (1-\Delta_{\rm eff}^2)}\right]
\end{equation}
Plugging this expression back in Eq.~\ref{eq:h2} and following with some algebra we get the sine-Gordon Hamiltonian with effective parameters:
\begin{equation}
{\mathbb h}={\mathbb h}_0+\frac{J_{\rm eff} S^2}{2}\left(\frac{d a}{d x}-\alpha\right)^2-\frac{h_{\rm eff}^2 \sin^2 a}{8 J_{\rm eff}},
\end{equation}
where
\begin{eqnarray}
{\mathbb h}_0 &=& -\tilde{J} S^2 -\frac{h_z^2}{4 \tilde{J} (1+\Delta_{\rm eff})}-\frac{h_z^2 h_x^2}{16 \tilde{J}^3 S^2 (1+\Delta_{\rm eff})(1-\Delta_{\rm eff}^2)}\nonumber\\
J_{\rm eff}&=&\tilde{J}\left[1-\frac{h_z^2}{4 \tilde{J}^2 S^2 (1+\Delta_{\rm eff})^2}\right]\nonumber\\
h_{\rm eff}&=& h_x\sqrt{\left[1-\frac{h_z^2}{2 \tilde{J}^2 S^2 (1+\Delta_{\rm eff})(1-\Delta_{\rm eff}^2)}\right]\left[1-\frac{h_z^2}{4 \tilde{J}^2 S^2 (1+\Delta_{\rm eff})^2}\right]}\
\nonumber\\
&\simeq & h_x\sqrt{1-\frac{h_z^2}{4 \tilde{J}^2 S^2 }\frac{3-\Delta_{\rm eff}}{(1+\Delta_{\rm eff})^2(1-\Delta_{\rm eff})}}
\end{eqnarray}
Following the same procedure as in Ref.~[\onlinecite{soliton}], the critical field for the commensurate-incommensurate transition is given by
\begin{equation}
h_{\rm eff}=\pi \alpha J_{\rm eff},
\end{equation}
which yields
\begin{equation}
h_{x,c}\left[1-\frac{h_z^2}{8 \tilde{J}^2 S^2 }\frac{3-\Delta_{\rm eff}}{(1+\Delta_{\rm eff})^2(1-\Delta_{\rm eff})}\right]\simeq\pi\alpha\tilde{J} S \left[1-\frac{h_z^2}{4 \tilde{J}^2 S^2 (1+\Delta_{\rm eff})^2}\right]
\end{equation}
After some quick algebra this results in
\begin{equation}
h_{x,c}=\pi \alpha \tilde{J} S \left[1+\frac{h_z^2}{8 \tilde{J}^2 S^2 (1-\Delta_{\rm eff}^2)}\right],
\end{equation}
which is precisely Eq.~(\ref{eq:hc2}).

\section{Case Studies: Simple Regions of the Quantum Phase Diagram}

The objective of this Appendix is to verify the consistency of Eqs. ~(\ref{eq:init}), ~(\ref{eq:rg_nab}) and Table I for a variety of cases in which Bethe ansatz solutions are available.
As a byproduct we derive an expression for the constant $c$ defined through $\lambda_{\rm xc}=c(1-\Delta)$, and comment on our choice for the RG cutoff energy scale.

(i) XXZ model with $h_x=h_z=0$ and $D\neq 0$.

In this case $\theta_R=0$, $\theta_L=-\pi$ and thus $y_A(0)=\tilde{y}_A(0)=0$.
It follows that $y_A(l)=\tilde{y}_A(l)=0$.
Accordingly the pertinent RG equations become
\begin{equation}
\label{eq:xxz}
\frac{d y_x}{d l} = y_z y_y \mbox{  ;   } \frac{d y_y}{d l} = y_z y_x \mbox{  ;   } \frac{d y_z}{d l}=y_x y_y,
\end{equation}
with initial conditions given by $y_x(0)=-y_y(0)=g_{\rm bs}/(2\pi v)$ and $y_z(0)=g_{\rm bs}(1+\lambda)/(2\pi v)$.
$y_x(0)=-y_y(0)$ implies $y_x(l)=-y_y(l)$ and thus there is no instability towards ``$\epsilon$'' or ``$N^x$''.
Moreover because $y_x(l)=-y_y(l)$ we are left with Eq.~(\ref{eq:kt_nab}) for $y_C$ and $y_\sigma$, which applies when $l<l_\phi$ as well as when $l>l_\phi$.
The analytical solutions of the Kosterlitz-Thouless equations dictate that when $\lambda>0$ (easy-plane anisotropy) the system flows to the gapless LL phase ($y_C\to 0$), whereas when $\lambda<0$ (easy-axis anisotropy) the system evolves to ``$N^z$'' ($y_C\to\infty$).
 We reiterate that the $O(D^2/J^2)$ term in the bosonized form of ${\cal V}$ (cf. Eq.~(\ref{eq:h_def}))  is crucial in order to get the correct answer for the case in which $\Delta=1$ and $D\neq 0$.
In particular, for $\lambda>0$ Bethe ansatz calculations prove that $y_C\to 0$ and $y_\sigma\to 2(1-K_{\rm inf})$, where $K_{\rm inf}^{-1}=1-\cos^{-1}(\Delta_{\rm eff})/\pi$. 
With this in mind we evaluate the value of the constant $c$ which enters the definition of $\lambda$.
We start by recognizing that Eq.~(\ref{eq:xxz}) implies $y_\sigma(0)^2-y_C(0)^2=y_\sigma(\infty)^2-y_C(\infty)^2$ with $y_C(0)=g_{\rm bs}/2\pi v$ and $y_\sigma(0)=-y_C(0)(1+\lambda)$. 
This results in $y_\sigma(\infty)\simeq -\sqrt{2\lambda} g_{\rm bs}/ 2\pi v\simeq -\sqrt{2 c}\sqrt{1-\Delta_{\rm eff}} g_{\rm bs}/2 \pi v$, where we have neglected $O(\lambda^2)$ and $O((1-\Delta) D^2/J^2)$ terms.
Comparing this with the Bethe ansatz prediction we obtain 
\begin{equation}
\label{eq:c}
c=\left(\frac{2}{\pi} \frac{2\pi v}{g_{\rm bs}}\right)^2\simeq 7.66.
\end{equation}

(ii) XXZ model with $h_x=D=0$ and $h_z\neq 0$. 

In this case $\theta_R=\theta_L=0$. 
Namely, the ``chiral'' rotation is simply the identity.
In this case too $y_A(l)=\tilde{y}_A(l)=0$ and the RG equations are given by Eq.~(\ref{eq:xxz}).
The initial conditions are $y_x(0)=y_y(0)=-g_{\rm bs}/(2\pi v)$ and $y_z(0)=y_y(0)(1+\lambda)$.
It follows that $y_C(l)=0$ and Eq.~(\ref{eq:xxz}) turns into KT equations for $y_B$ and $y_\sigma$. 
For $\lambda<0$ the system flows to $y_B(l)\to -\infty$, which in the original coordinates corresponds to ``$N^z$'' (recall Eq.~\ref{eq:op_transf}).
$y_B$ reaches strong coupling ($y_B\equiv -1$) when $l=l_c=(2\pi v/g_{\rm bs}) (\pi-\cos^{-1}(1+\lambda))/\sqrt{-2\lambda}$ for $\lambda\lesssim 0$.
However, $h_z$ interrupts the flow of $y_B$ at $l=l_\phi$ (note that $t_\theta=0$) and thus ``$N^z$'' is the ground state only when $l_\phi>l_c$; for $l_\phi<l_c$ the ground state is $LL$.
Reaching strong coupling requires $h_z<h_c$ where 
\begin{equation}
\label{eq:h_zc}
h_c\propto T_0\exp\left[-\frac{2\pi v}{g_{\rm bs}}\frac{\pi}{\sqrt{2 c (\Delta-1)}}\right]
\end{equation}
is the critical field defined through $l_\phi=l_c$.
The prefactor in Eq.~(\ref{eq:h_zc}) is somewhat arbitrary because it depends on the precise value of $y_B(l)$ for which one decides that ``strong coupling'' has been reached.
The critical field is also known from Bethe ansatz calculations,\cite{cabra2004} which dictate $h_c/J\propto \exp[-\pi^2/2\sqrt{2(\Delta-1)}]$ for $\Delta\gtrsim 1$.
Matching the exponent of this expression with that of Eq.~(\ref{eq:h_zc}) yields $c=7.66$, in agreement with Eq.~(\ref{eq:c}).
On the other hand, for $\lambda>0$ Eq.~(\ref{eq:xxz}) flows to a weak-coupling regime regardless of $h_z$ (provided that $h_z<<J$).  
Once again this $LL$ phase agrees with Bethe ansatz predictions.

(iii) XXZ model with $h_z=D=0$ and $h_x\neq 0$. 

In this case $\theta_R=\theta_L=-\pi/2$ and thus $y_A(l)=\tilde{y}_A(l)=0$.
The flow equations are once again given by Eq.~(\ref{eq:xxz}), with 
$y_y(0)=y_z(0)=-g_{\rm bs}/(2\pi v)$ and $y_x(0)=-g_{\rm bs}(1+\lambda)/(2\pi v)$.
These RG equations must be replaced by Eq.~(\ref{eq:kt_nab}) at $l\geq l_\phi$.
We find that for $\lambda>0$ the ground state is ``$N^y$'' and while for $\lambda<0$ the system flows to ``$N^z$''.
These results are in concordance with the classical considerations of Section I and agree with independent quantum mechanical calculations.\cite{xxz transverse}
We note in passing that there is no $LL$ phase in the XXZ model with a {\em uniform} transverse field (i.e. $D=0$ in Eq.~(\ref{eq:h2b})).

(iv) XXX model with $D=h_x=0$ and $h_z\neq 0$.
 
In this case $y_x(0)=y_y(0)=y_z(0)=-g_{\rm bs}/2\pi v$ and the stage I flow equations reduce to $d y_i/d l=y_i^2$ (for $i=x,y,z$), whose solution is
$y_i(l)=-(g_{\rm bs}/2\pi)/(1+l g_{\rm bs}/2\pi v)$.
All couplings stop renormalizing at $l_\phi=\log(v/a_0 h_z)$.
The Luttinger parameter $K=1-\frac{1}{2}y_\sigma(l_\phi)$ is then given by
\begin{equation}
\label{eq:K_LL}
K=1-\frac{1}{2\log\left(\frac{h_0}{h_z}\right)} \mbox{    ;    } h_0=\frac{v}{a_0 h_z} \exp\left(\frac{g_{\rm bs}}{2\pi v}\right).
\end{equation}
Using $T_0=v/a_0=0.077 J$ and $g_{\rm bs}=0.23\times (2\pi v)$ we get $h_0=5.95 J$.
This is slightly smaller than $h_0=J\sqrt{8\pi^3/e}=9.55 J$, obtained from solving the Bethe ansatz equations.\cite{korepin1993}
However, this discrepancy is masked by the fact that the precise value of $l_\phi$ is uncertain (we could have used $l_\phi=\log(A v/a_0 h_z)$, where $A$ is any constant of order one).

\end{widetext}


\begin{thebibliography}{50}
\bibitem{l_c} $l_c\simeq 10$ is found for $D=0.1 J$, $h_x\simeq 1.5 D$ and $\Delta\simeq 1$. For smaller $D$ and/or $\Delta$, $l_c$ increases rapidly.
\end{thebibliography}
\end{document}